\documentclass[aps,prx,twocolumn,superscriptaddress,showpacs,longbibliography,floatfix]{revtex4-2}
\usepackage{xcolor}
\usepackage{graphicx}
\usepackage{amsmath}
\usepackage{amssymb}
\usepackage{bm}
\usepackage{algorithm}
\usepackage{algpseudocode}
\usepackage{multirow}
\usepackage{physics}
\usepackage{hyperref}
\usepackage{booktabs}
\usepackage{placeins}

\begin{document}
\newcounter{para}
\newcommand{\para}[1]{\par\refstepcounter{para}\textbf{{\color{cyan}[\thepara]}}\space\textbf{\textcolor{cyan}{#1}}}

\newcommand{\KA}[1]{\textcolor{purple}{KA:#1}}
\newcommand{\HP}[1]{\textcolor{cyan}{HP:#1}}
\newcommand{\HPadd}[1]{\textcolor{cyan}{#1}}
\newcommand{\HPcom}[1]{\textcolor{blue}{[HP:#1]}}

\newcommand{\hl}[1]{\textcolor{red}{#1}}
\newcommand{\KAcom}[1]{\textcolor{purple}{[KA:#1]}}

\newcommand{\jp}[1]{\textcolor{blue}{#1}}
\newcommand{\mg}[1]{\textcolor{magenta}{#1}}

\begin{abstract}
We develop classical simulation algorithms for monitored quantum circuits that produce states with low levels of ``magic'' (i.e., non-stabilizerness).  These algorithms are particularly well-suited to circuits with high rates of Pauli measurements, such as those encountered in quantum error correction and monitored quantum circuits.  The measurements serve to limit the buildup of magic induced by non-Clifford operations arising from generic noise processes or unitary gates, respectively. Our algorithms also allow a systematic truncation procedure to achieve approximate simulation.  To benchmark our approach, we study the dynamics of all-to-all monitored quantum circuits with a sub-extensive rate of $T$-gates per unit of circuit depth, where we can simulate previously inaccessible system sizes and depths. We characterize measurement-induced phase transitions in the output wavefunction, including in the entanglement, purification, and magic. We outline the utility of our algorithms to simulate dynamics with low magic and high entanglement, complementary to the leading matrix-product state approaches.
\end{abstract}

\title{Classical Simulations of  Low Magic Quantum Dynamics}

\author{Kemal Aziz}
 \affiliation{Department of Physics and Astronomy, Center for Materials Theory, Rutgers University, Piscataway, NJ 08854, USA}
\author{Haining Pan}
 \affiliation{Department of Physics and Astronomy, Center for Materials Theory, Rutgers University, Piscataway, NJ 08854, USA}
\author{Michael J. Gullans}
 \affiliation{Joint Center for Quantum Information and Computer Science, NIST/University of Maryland, College Park, Maryland 20742, USA}
\author{J. H. Pixley}
 \affiliation{Department of Physics and Astronomy, Center for Materials Theory, Rutgers University, Piscataway, NJ 08854, USA}
 \affiliation{Center for Computational Quantum Physics, Flatiron Institute, 162 5th Avenue, New York, NY 10010, USA}

\maketitle

\section{Introduction}

Simulating the dynamics of quantum many-body systems is a fundamental challenge, as the Hilbert space grows exponentially with system size, rendering exact state-vector methods intractable for all but the smallest systems. Although tensor network methods such as matrix product states (MPS) can efficiently simulate states with area-law entanglement, they face exponential scaling costs when confronted with the volume-law entanglement characteristic of generic quantum dynamics~\cite{schollwock2011density,vidal2004efficient}.

Therefore, developing new classical simulation algorithms is crucial for advancing our understanding of key areas like quantum error correction (QEC) and measurement-induced phase transitions (MIPTs).
In QEC, robust simulations are necessary to model fault tolerance in noisy, highly entangled circuits~\cite{fowler2012surface, terhal2015quantum}, while in MIPTs, large-scale simulations are essential for characterizing universal scaling behavior~\cite{li2018quantum, Skinner2019Measurement, choi2020quantum,Zabalo2020critical}.
A powerful tool for studying highly entangled dynamics is the stabilizer formalism, which allows for the efficient simulation of Clifford circuits~\cite{gottesman1998heisenberg,aaronson2004improved}.
However, Clifford circuits are not universal and cannot capture the generic quantum phenomena that arise from non-Clifford operations~\cite{Bravyi2005Universal, Bravyi2016Improved}.
Since simulation complexity grows exponentially with the number of non-Clifford gates, there is a strong motivation to develop methods that can efficiently handle circuits that are ``close" to the stabilizer manifold~\cite{bravyi2019simulation, gonzalez2025pauli, Bennink2017Unbiased}.

The amount of non-Clifford resources required to prepare a state is known as non-stabilizerness, commonly referred to as ``magic''. Magic is zero for stabilizer states and increases monotonically under the application of non-Clifford operations. In recent years, a range of near-stabilizer simulation methods have been developed to simulate quantum circuits with limited non-Clifford resources efficiently. These include Monte Carlo techniques based on stabilizer decompositions~\cite{Bravyi2005Universal,Schwarz2013Simulating,Bejan2024Dynamical}, mapping to classical statistical mechanics models~\cite{Tarabunga2025Magic} as well as tensor network methods such as MPS simulations~\cite{Fux2024Entanglement} and Clifford-augmented Matrix Product States (CAMPS)~\cite{Liu2025Classical, Qian2024Augmenting}. Additionally, hybrid approaches such as stabilizer tensor networks, which combine stabilizer formalism with tensor network contraction and sampling, have been developed ~\cite{MasotLlima2024Stabilizer}. Collectively, these techniques enable approximate classical simulation of quantum circuits that are not entirely Clifford but remain structured enough to avoid worst-case exponential scaling of simulation complexity with the number of non-Clifford gates.

In the following manuscript, we avoid the sampling overhead of existing near-stabilizer techniques by developing a near-Clifford simulator that uses an exact representation of the state. Instead of sampling over Clifford circuits, our approach updates the stabilizer decomposition dynamically during the circuit evolution. Our approach is designed for dynamic measurement and mid-circuit branching, such as those encountered in quantum error correction and MIPTs~\cite{Skinner2019Measurement,Li2019measurement}. Existing stabilizer rank techniques do not generalize easily to dynamic, measurement-driven evolutions.
Our approach can also simulate states with volume law entanglement, for a low number of $T$-gates, unlike MPS approaches.
The Clifford+$T$ gate set constitutes the minimal extension of stabilizer circuits to universality, making it the natural and minimal framework for studying the onset of non-stabilizer physics in monitored quantum circuits.

To benchmark our algorithm in a non-trivial setting, we consider random circuit models with two-qubit gates and single-site measurements, but the gates are allowed to act on arbitrary pairs of qubits in the system that are chosen at random.   Such ``all-to-all'' models can exhibit MIPTs, but the usual notion of an area-to-volume law entanglement transition that arises in spatially local models is revised~\cite{Nahum2021Measurement}. These models are particularly difficult for MPS-based simulations due to the lack of locality and are therefore useful as a general test case for our purposes.

The remainder of this paper is organized as follows.  In Section \ref{sec:CircuitSimulation}, we introduce our algorithm for updating near-stabilizer states under the action of $T$-gates. We also define the measure of magic, stabilizer nullity, which we study in this work. In Section \ref{sec:CircuitProtocol}, we introduce the models of monitored circuits, including entangling gates and protocols for the measurements. In Section \ref{sec:magic_complexity}, we test the accuracy of our algorithm, which simulates the full density matrix, by varying a cutoff parameter that controls the accuracy of the algorithm. In Section \ref{sec:space_complexity}, we examine the space complexity of these circuits, namely how the amount of memory to store the state scales with the density of $T$-gates. In Section \ref{sec:MIPT_transitions}, we study the MIPTs in these models, focusing on the magic and purification transitions. In Appendix \ref{sec:algorithms}, we provide more details on the algorithms used to study these circuits. These include the evolution of the density matrix of the generic state $\ket{\psi}$, $\rho=\ketbra{\psi}$, under measurements and Clifford unitaries, computing the R\'enyi entanglement entropies, and Bell sampling.
In Appendix \ref{sec:EntanglementTransitions}, we examine entanglement transitions in more detail. Namely, we locate cluster and all-to-all entanglement transitions. We also provide more details on locating the purification transition.  In Appendix \ref{sec:magic_analysis}, we include details on the procedure used to locate the magic transition.

\section{Near stabilizer simulation algorithm}
\label{sec:CircuitSimulation}
To set the stage, we begin by reviewing the stabilizer formalism that we will be perturbing with a dilute amount of $T$-gates.
\subsection{Stabilizer Formalism}
The density matrix of a general mixed stabilizer state on $L$ qubits can be written as a sum over the elements of its stabilizer group $\mathcal{S}$ \cite{aaronson2004improved}:
\begin{equation}
\rho_{S} = 2^{r-L} \prod_{i=1}^{r} \left( \frac{I + g_{i}}{2} \right) = \frac{1}{2^{L}} \sum_{g \in \mathcal{S}} g.
\label{eq:stabilizerstate}
\end{equation}
Here, $\mathcal{S}$ is an abelian subgroup of the $L$-qubit Pauli group $\mathcal{P}_L$ that does not contain $-I$. The generators $g_{i}$ of $\mathcal{S}$ are tensor products of Pauli matrices ($X$, $Y$, $Z$),
\begin{equation}
g_{i} = \bigotimes_{j=1}^{L} P_{ij}, \qquad P_{ij}\in\{I, X, Y, Z\}.
\end{equation}
An operator $g \in \mathcal{S}$ is said to stabilize $\rho_{S}$ if $g \rho_{S} g^{\dagger} = \rho_{S}.$ For a set of $r$ independent generators, the stabilized subspace has dimension $2^{L-r}$. Pure stabilizer states correspond to a full set of $r=L$ generators, while mixed stabilizer states have $r < L$.

\begin{figure*}[htbp]
    \centering
    \includegraphics[width=7in]{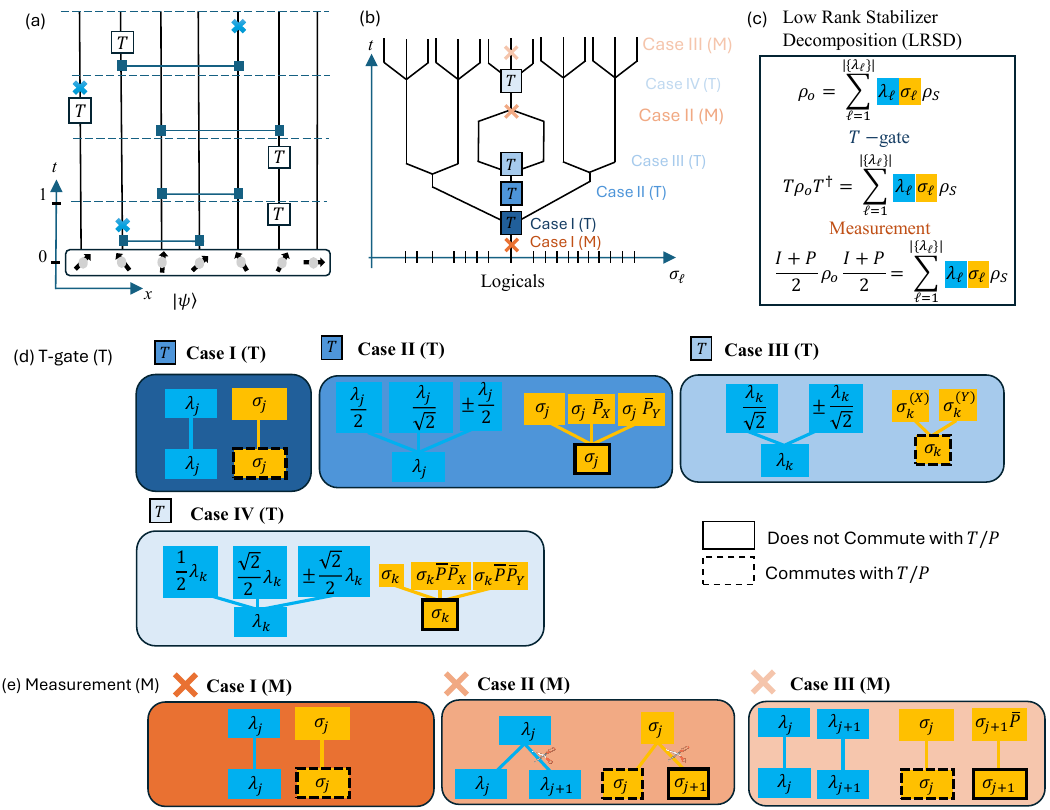}
    \caption{
(a) Circuit diagram for the evolution of the initial state that corresponds to the initial density matrix $\rho_o=\sum_{\ell} \lambda_{\ell} \sigma_{\ell} \rho_{S}$ with measurements, $T$-gates and Clifford unitaries (b) Schematic of the forking of the logicals $\sigma_{\ell}$ for the initial state under the circuit evolution in (a). The LRSD is represented diagrammatically, where each node represents $\sigma_{\ell}$ or $\lambda_{\ell}.$ Vertical edges connect terms from the initial state $\rho_o$ to the final state after the $T$-gate $T \rho_o T^{\dagger}$, or measurement of Pauli operator $P$, $\frac{I+P}{2} \rho_o \frac{I+P}{2}$. Each $\sigma_{\ell}$ is mapped to a new set of $\sigma_{\ell}$ attached to it by vertical lines. An X, or box with T, on a branch denotes measurement, or $T$-gate, respectively, in the entire circuit. (c) Equation for the LRSD and the effect of $T$-gates and measurements. (d) Each injection of a $T$-gate can fork each logical onto itself (Case I), all logicals into three new logicals (Cases II, IV), or certain logicals to two new logicals (Case III). The Pauli operator $\overline{P}$ is defined in Eq.~\eqref{eq:psi_decomposition}. The Pauli $\overline{P}_{(X/Y)}$ is defined by $\overline{P}_{X/Y}=\overline{P}_{1} \otimes \cdots \otimes (X/Y) \otimes \cdots \otimes \overline{P}_{L},$ where $X/Y$ and the $T$-gate is on the $i^{\mathrm{th}}$ qubit and likewise $\sigma_{\ell}^{(X/Y)}=\sigma_{\ell_{1}} \otimes \cdots \otimes (X/Y) \otimes \cdots \otimes \sigma_{\ell_{L}}.$ (e) Each measurement can fork each logical onto itself (Case I), eliminate all anticommuting logicals (Cases II), or update anticommuting logicals with the Pauli operator $\overline{P}$. Algorithm~\ref{alg:msm_near_clifford} in Appendix~\ref{sec:algorithms} provides details on measurement updates.
}
\label{fig:circuit_schmetic}
\end{figure*}

\subsection{Non-Clifford Evolution}
\label{sec:non-clifford-evolve}
Given an arbitrary pure quantum state $\ket{\psi}$, the stabilizer group of $\ket{\psi}$ is the set of Pauli operators that have eigenvalue one on that state
\begin{equation}
\label{eq:StabilizerGroup}
\mathcal{S}(\ket{\psi}) = \{P \in \mathcal{P}_L \mid P\ket{\psi} = \ket{\psi}\},
\end{equation}
where $\mathcal{P}_L$ is the Pauli group on $L$ qubits and $\mathcal{S}$ is a subgroup of $\mathcal{P}_L$.
We define the low-rank stabilizer decomposition (LRSD) of $\ket{\psi}$ as the representation
\begin{equation} \label{eq:LSRD_decomposition}
\ketbra{\psi} = \sum_{l \in \mathcal{L}_{\mathrm{LRSD}}} \lambda_l \sigma_l \rho_S.
\end{equation}
Here, $\rho_{S}$ is a (possibly mixed) stabilizer state that is defined solely in terms of $S$ through Eq.~\eqref{eq:stabilizerstate}. The coefficients $\lambda_{l} = \mel{\psi}{\sigma_{l}}{\psi}$ are real numbers and the ``logical'' operators $\sigma_{l}$ are elements of the Pauli group that commute with $S$. The set of all logical operators modulo phases $\mathcal{L}_{\mathrm{LRSD}}$ has size $4^{k}$ for a stabilizer group of size $2^{L-k}$ \cite{aaronson2004improved}.
 If $\rho_S$ is pure, then $\ket{\psi}$ contains no magic, and Eq.~\eqref{eq:LSRD_decomposition} consists of a single term: specifically, $\lambda_l = 1$ and $\sigma_l = I$. In this case, $\ket{\psi}$ is a stabilizer state. More generally, for a broad class of near-stabilizer states with small $k$, the number of logical operators needed in the LRSD remains small, allowing for an efficient classical representation.  This approach isolates the non-stabilizer content of the state into a compact, low-dimensional structure, the logical operator set, while leveraging the efficient classical simulability of stabilizer states. This representation builds on the stabilizer rank framework~\cite{bravyi2019simulation, Bravyi2016Improved}, which decomposes near-Clifford states into sums of stabilizer states. The specific structured form used here, expressing a state in terms of the logical operators of a single stabilizer code, was introduced in Ref.~\cite{Hangleiter2024bell}. We note that while Eq.~\eqref{eq:LSRD_decomposition} is written for the pure-state projector $\ketbra{\psi}$ of a single measurement trajectory, the LRSD naturally generalizes to mixed states $\rho = \sum_l \lambda_l \sigma_l \rho_S$ with real coefficients $\lambda_l$.

We now present how to update the LRSD representation of the state under the action of $T$-gates. For a $T$-gate on the $i^{th}$ qubit, $T_{i}$, the state evolves as:
\begin{equation}
\ketbra{\psi} \rightarrow T_{i} \ketbra{\psi} T_{i}^{\dagger} = \sum_{l \in \mathcal{L}_{\mathrm{LRSD}}} \lambda_{l} T_{i} \sigma_{l} \rho_{S} T_{i}^{\dagger}.
\end{equation}
Up to an overall phase, the $T$-gate can be rewritten in the following form:
\begin{equation}
T_{i} = \alpha I^{\otimes L} + \beta P,
\qquad
P = I^{\otimes (i-1)} \otimes Z \otimes I^{\otimes (L-i)}
\end{equation}
Here, $\alpha = \cos\left(\frac{\pi}{8}\right)$ and $\beta = -i\sin\left(\frac{\pi}{8}\right)$. The overall phase is irrelevant because it is canceled when the density matrix is acted upon by conjugation. This form of the decomposition is useful because it separates the action of the $T$-gate into parts that either commute or anticommute with Pauli operators. This simplifies the update rules. In each update with $T_{i}$, the complexity of the LRSD will depend on the commutation relations between $P$ with the stabilizers and logicals of $\rho_{S}.$
Specifically, there are four cases to consider when evolving the state under $T$-gates. In the following, we use $\mathcal{S}$ to denote the stabilizer group of $\rho_{S}$, and $[P,\mathcal{S}]=0$ to mean that $P$ commutes with every element of $\mathcal{S}$ and likewise for $\mathcal{L}_{\mathrm{LRSD}}$.

\begin{itemize}
    \item \textbf{Case I:} If $P \in \mathcal{S}$, then $[P, \mathcal{L}_{\mathrm{LRSD}}] = 0$ and $[P, \mathcal{S}] = 0$.
    In this scenario, $\rho$ remains unchanged.

    \item \textbf{Case II:} If $[P, \mathcal{S}] \neq 0$ but $[P, \mathcal{L}_{\mathrm{LRSD}}] = 0$, then
    \begin{equation}
    \label{eq:UpdateWt}
    \rho \rightarrow \sum_{l \in \mathcal{L}_{\mathrm{LRSD}}} \lambda_{l} \sigma_{l} T_{i} \rho_{S} T_{i}^{\dagger}.
    \end{equation}
    We perform the stabilizer decomposition using Algorithm \ref{alg:decomposition}:
    \begin{equation}
    \label{eq:psi_decomposition}
    \rho_{S} = \frac{1 + \bar{P}}{2}\rho_{S}^{P},
    \end{equation}
    where $[P, \rho_{S}^{P}] = 0$ . The action of the $T$-gate on the Pauli string $\bar{P}$ is given by
    \begin{equation}
    T_{i} \frac{1 + \bar{P}}{2} T^{\dagger}_{i},
    \end{equation}
    which can be deduced from the rule:
    \begin{equation}
    \label{eq:T_Update}
    T_{i} \sigma_{q} T_{i}^{\dagger} = (\alpha^{2} + \beta^{2}) \sigma_{q} + 2i\alpha\beta \epsilon_{qr3} \sigma_{r}.
    \end{equation}
    Here, $\epsilon_{qr3}$ is the Levi-Civita symbol and $\sigma_{q}$ is a single-qubit Pauli matrix, where each index runs from $1$ to $3$.
    \item \textbf{Case III:} If $[P, \mathcal{S}] = 0$ but $[P, \mathcal{L}_{\mathrm{LRSD}}] \neq 0$, then $P$ commutes with all the stabilizers but is not itself in the stabilizer group.
    In this case:
    \begin{equation}
    \rho = \sum_{l \in \mathcal{L}_{\mathrm{LRSD}}} \lambda_{l} T_{i} \sigma_{l} T_{i}^{\dagger} \rho_{S}.
    \end{equation}
    Each term in the tensor product of $\sigma_{l}$ is updated using Eq.~\eqref{eq:T_Update}.

    \item \textbf{Case IV:} If $[P, \mathcal{L}_{\mathrm{LRSD}}] \neq 0$ and $[P, \mathcal{S}] \neq 0$, the anticommuting logicals can be updated so that $[P, \mathcal{L}_{\mathrm{LRSD}}] = 0$.
    If $[\sigma_{l}, P] \neq 0$, then
    \begin{equation}
    \sigma_{l} \rightarrow \sigma_{l} \bar{P}.
    \end{equation}
    This transformation leaves the state invariant as $\bar{P} \rho_{S} = \rho_{S}, \forall \bar{P} \in \mathcal{S}$. After this update, $[P,\mathcal{L}_{\mathrm{LRSD}}]=0$, which reduces to Case II.
\end{itemize}

The treatment of Clifford unitaries and Pauli measurements appears in Appendix \ref{sec:algorithms}.
We can now provide some intuition for how the complexity of the classical simulation evolves. As visually summarized in Fig. 1, each $T$-gate application branches the logical operators into new paths, while subsequent measurements can collapse or eliminate them. Each $T$-gate application can increase the number of logicals $|\mathcal{L}_{\mathrm{LRSD}}|$. Specifically, in Cases II and IV, the number of logicals can triple, while in Case III, they can double. Conversely, each measurement can reduce the number of logicals, depending on the commutation relationships. If $[P, \mathcal{L}_{\mathrm{LRSD}}] = 0$, or if $[P, \mathcal{L}_{\mathrm{LRSD}}] \neq 0$ and $[P, \mathcal{S}] \neq 0$, the number of logicals remains the same. However, if $[P, \mathcal{L}_{\mathrm{LRSD}}] \neq 0$ and $[P, \mathcal{S}] = 0$, the number of logicals can decrease by up to a factor of three. The detailed procedure is provided in Algorithm~\ref{alg:decomposition}.

Introducing $T$-gates takes us out of the stabilizer manifold. Here, we quantify the non-stabilizerness of a state through its ``distance'' from the set of stabilizer states~\cite{Bravyi2005Universal,Haug2023scalable,Leone2022stabilizer} and is commonly referred to as ``magic''.
For a pure stabilizer state, the magic is by definition zero. For a non-stabilizer state, the magic quantifies the degree to which the state deviates from the stabilizer set, and serves as a resource-theoretic measure of the non-Clifford operations required for its preparation. While high magic often correlates with greater circuit complexity \cite{Bravyi2016Improved}, the experimental difficulty of preparing a given non-stabilizer state may vary depending on the physical platform and available operations\cite{Bermejo-Vega2018Architectures,Wang2024Efficient}.

We now define a measure of magic called the stabilizer nullity~\cite{Beverland2020Lower}. If $\ket{\psi}$ is an $L$ qubit state, the stabilizer group $\mathcal{S}$, of $\ket{\psi}$ is defined in Eq.~\eqref{eq:StabilizerGroup}
The stabilizer nullity of $\ket{\psi}$, $\mathcal{M}(\ket{\psi})$, is
\begin{equation}
\label{eq:magic_equation_main}
\mathcal{M}(\ket{\psi}) = L - \mathrm{log}_{2}\abs{\mathcal{S}(\ket{\psi})},
\end{equation}
where $\abs{\mathcal{S}(\ket{\psi})}$ is the size of the stabilizers.
For a stabilizer state, $\mathcal{M}=0.$ For non-stabilizer states, $\mathcal{M}>0.$ The stabilizer nullity measures the number of non-stabilizer degrees of freedom in the state. It serves as an efficiently computable upper bound on more general measures of magic, such as stabilizer rank or robustness of magic \cite{Bravyi2016Improved, howard2017application, seddon2019quantifying, SuzukiQuantum2022}.

\section{Circuit Model}
\label{sec:CircuitProtocol}
\begin{figure}[htbp]
    \includegraphics[width=3.4in]{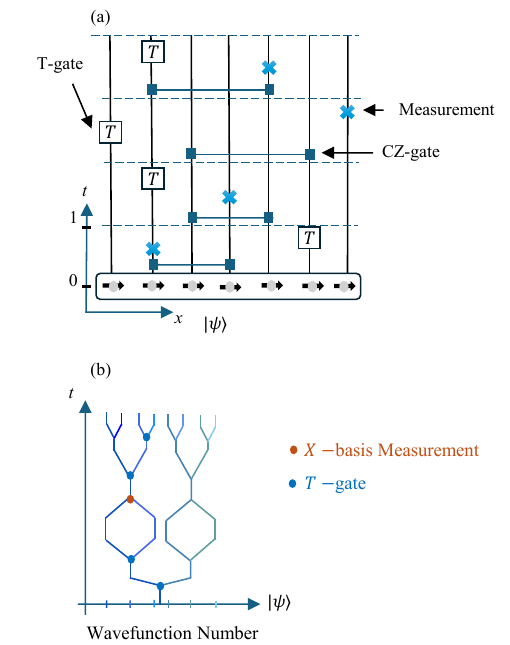}
    \caption{
    (a) Circuit diagram for the single-pair all-to-all model. Each time step consists of a CZ gate applied between two random sites, followed by a measurement on a randomly chosen site in either the $X$ or $Z$ basis. Each dashed line represents the end of a single time step in the model. (b) Forking of the initial stabilizer wavefunction $\ket{\psi}=\ket{+}^{\otimes L}$ under circuit evolution with $T$-gates, measurements, and Clifford unitaries. Each injection of $T$-gate takes a stabilizer state $\ket{\psi}$ onto a superposition of at most two stabilizer states $\ket{\psi'}=c_{+}\ket{\psi_{+}}+c_{-}\ket{\psi_{-}}.$ Each branch represents a different stabilizer state. Measurements can collapse different branches onto a single branch. Each quantum trajectory is represented by a different color.
    }
    \label{fig:circuit-schematic}
\end{figure}
In all-to-all random circuit models, there is no constraint of spatial locality. Part of the motivation of studying these models is that a graphical description of the MIPT in these circuits leads to a precise connection between the MIPT and mean-field percolation \cite{Vijay2020measurement}. This is analogous to the connection between two-dimensional classical percolation and the measurement-driven entanglement transition in one spatial dimension \cite{Skinner2019Measurement}. Furthermore, all-to-all models are directly relevant to various experimental platforms. For example, proposals for scaling up quantum computers often involve quantum network architectures, where locally interacting nodes are connected via non-local photonic interconnects, leading to effectively all-to-all connectivity at large scales \cite{Kimble2008Quantum}. At the hardware level, trapped ion quantum computers already realize all-to-all gate connectivity among physical qubits at intermediate scales \cite{Noel2022Observation,Monroe2014Large}. These experimental developments further underscore the importance of understanding the fundamental features of MIPTs in nonlocal circuit architectures.
We now introduce the circuit protocols considered in this paper. We focus on an all-to-all geometry where the two-site unitary can be applied to any pair of qubits, which is an effective 0D system. Specifically, we consider single-pair all-to-all models, where in a single time step, at most one gate couples a pair of qubits, as shown in Fig.~\ref{fig:circuit-schematic}.
Each time step consists of three probabilistic stages: Clifford unitaries, non-Clifford unitaries, and measurements.

The single-pair all-to-all model, as shown in Fig.~\ref{fig:circuit-schematic}, is adapted from Ref.~\cite{Vijay2020measurement}.
In the first stage, we apply a control-$Z$ (CZ) gate, as defined in Eq.~\eqref{eq:CZ-gate} with a probability of $1/2$ to a specific pair of qubits~\footnote{This choice ensures that the circuit is in a nontrivial entangling regime: choosing the probability of a CZ gate arbitrarily as $p_{u}$, if $p_u$ is too small it drives the system into a trivial area-law entangled phase for all non-zero measurement rates.}. This pair is chosen uniformly from all combinations of pairs.
After the Clifford unitary gate, we apply a $T$-gate with a probability of $p_T$ to a randomly chosen qubit. Note that $p_T$ is therefore a per-timestep rate, not a per-qubit rate.
The $T$-gate is the simplest non-Clifford gate that, combined with Clifford operations, promotes the gate set to universality. Our model therefore represents the minimal circuit architecture needed to capture the departure from the stabilizer manifold.
Finally, we perform a projective local measurement with a finite probability onto a specific qubit chosen randomly.

In particular, we study two distinct models of quantum circuits that are distinguished by the basis in which projective measurements are performed: the $Z$ or $X$ basis. We denote the probability to make a measurement along the $z$-axis as $p_{m,Z}$ and the $x$-axis as $p_{m,X}$. As with $p_T$, these are per-timestep rates, not per-qubit rates.
In the $Z$ basis model, measurements are diagonal in the computational basis, which allows for an efficient classical simulation for all rates of $T$-gates to remain. However, computing the amount of non-stabilizerness (i.e., the magic) of these states remains a computational challenge, and one of the algorithms we develop allows for its efficient computation (technical details are in Appendix \ref{sec:BellSample}). In contrast, the $X$-basis measurements act in a rotated basis, which can generate additional entanglement and simulation difficulty. Our newly developed algorithm (in Appendix \ref{sec:clifford_algorithms}) is able to still efficiently simulate the circuit evolution in the presence of a sub-extensive set of $T$-gates (i.e. $p_T\sim 1/L$ or $\sim 1/L^2$).

The entangling unitary in our model is the CZ gate in the Clifford group, acting on two qubits $i$ and $j$ as
\begin{equation}
\text{CZ}_{ij}\ket{s_{i} s_{j}} = (-1)^{s_{i}  s_{j}}\ket{s_{i} s_{j}},
\label{eq:CZ-gate}
\end{equation}
where $s_i, s_j \in \{0,1\}$ are the computational basis states.
The CZ gate is a Clifford gate, meaning that it maps stabilizer states onto stabilizer states. That is, the set of stabilizing Pauli operators for the output state can be obtained by conjugating the stabilizers of the input state with the CZ gate. This transformation preserves the stabilizer structure and can be tracked efficiently using stabilizer formalism.

We consider a single-qubit projective measurement $M_{\alpha,\pm}^{(j)}$ along the $ \alpha$ direction in the mid-circuit as
\begin{equation}
\label{eq:clifford_measurement}
M_{\alpha,\pm}^{(j)}\ket{\psi} = \frac{\Pi_{\alpha, \pm}^{(j)} \ket{\psi}}{\left\| \Pi_{\alpha, \pm}^{(j)} \ket{\psi} \right\|},
\end{equation}
where $\alpha$ is the axis of the measurement, and $\Pi_{\alpha,\pm}^{(j)} = \ketbra{\pm}_\alpha$ project qubit $j$ to the $\pm \alpha$ axis.
Here, we consider two bases of measurements, either along $z$ axis ($\alpha=z$), or along $x$ axis ($\alpha=x$), respectively.
For the measurement along the $z$-axis, we additionally rotate the measurement outcome for each qubit from $\ket{\pm z}$ to $\ket{+x}$ to keep them entangled under the unitary gates. Our architecture is adapted from the solvable circuit model introduced in \cite{Vijay2020measurement}, which demonstrates a novel phase transition between a fully-entangled phase and a separable phase composed of finite subsystems.

The second type of unitary gate is the single-qubit $T$-gate, which introduces a non-Clifford $\pi/4$ phase to the $\ket{-z}$, defined as
\begin{equation}
T = \begin{pmatrix} 1 & 0 \\ 0 & e^{i \pi / 4} \end{pmatrix}.
\end{equation}

Applying a $T$-gate to a stabilizer state $\ket{\psi}$ results in a superposition of at most two stabilizer states, $\ket{\psi'} = c_+ \ket{\psi_+} + c_- \ket{\psi_-}$ as depicted in Fig.~\ref{fig:circuit-schematic}. If $T$ anticommutes with at least one element of the stabilizer group, the stabilizer state is taken out of the stabilizer manifold. This branching increases the complexity of near-stabilizer simulators, which use the stabilizer formalism. A generic pure quantum state $\ket{\psi}$ can be written in terms of stabilizer states,
\begin{equation}
\ket{\psi} = \sum_{i=1}^{\chi} c_{i}\ket{\phi_{i}},
\end{equation}
where the stabilizer rank is $\chi(\ket{\psi})$, $c_{i} \in \mathbb{C}$, and $\ket{\phi_{i}}$ is a stabilizer state. The minimum number $\chi$ for which such a decomposition exists is the stabilizer rank.
\section{Complexity of Simulation}
\label{sec:magic}
\subsection{Simulation Task}
\label{sec:simulation_task}
The computational task studied in this work is the trajectory-resolved simulation of monitored quantum circuits. Specifically, for a single measurement trajectory of depth $t \sim L^{2}$, we store and evolve an explicit representation of the full density matrix
\begin{equation}
\rho(t) = \ketbra{\psi(t)}
\end{equation}
under probabilistic unitaries (CZ), $T$-gates, and projective measurements in either the X or Z basis. The goal is to maintain an exact truncation-controlled representation of $\rho(t)$, thus enabling computation of nonlinear observables such as entanglement entropies, purification times, and stabilizer nullity.

This is the number of entries, or bits and complex numbers, required to specify the state. The stabilizer state is represented by a tableau of size $(2L+1)\times(2L+1)$. Each Pauli operator $\sigma_{l}$ is represented by a $2L$ binary vector, as defined in Section \ref{sec:magic}. Each phase $\lambda_{l}$ is a complex number. Each logical Pauli operator $\sigma_{l}$ is specified by an $L$-component vector with entries in $\{0,1,2,3\}$, labeling $I,X,Y$, and $Z$, respectively.

These considerations mean that
\begin{equation}
\label{eq:NumberEntries}
    \mathcal{N} = (2L+1)^{2} +  \left|\{\lambda_l\}\right|L + \left|\{\lambda_l\}\right|
\end{equation}
entries are required to specify the state. The $(2L+1)^{2}$ factor is the number of entries required to specify $\rho_{S}.$ The $|\{\lambda_{l}\}|L$ factor is the number of entries required to store $\sigma_{l}.$ The $|\{\lambda_{l}\}|$ is the number of entries required to store the complex phases. Although the Pauli indexing is not unique, one could also use a length-$2L$ binary vector to specify $\sigma_{l}$, for example. However, all reasonable choices should give the same asymptotic scaling $\sim L^{2} + \left|\{\lambda_l\}\right|L + \left|\{\lambda_l\}\right|$. By contrast, a full state-vector description demands $2^{L}$ complex amplitudes, or complex double-precision numbers. Moreover, the state-vector stores exclusively complex numbers, whereas our representation mixes complex numbers, integers, and Booleans, so the count above is a conservative estimate of the true memory advantage of our approach.

We say that the circuit ensemble is efficiently simulable if, for trajectories drawn from the ensemble, the scaling of $\mathcal{N}(L)$ with system size $L$ is polynomial rather than exponential. The parameter regimes in which this condition is satisfied, along with numerical evidence and a detailed analysis of worst-case versus average-case scaling, are presented in Sec.~\ref{sec:space_complexity}.

\subsection{Magic Complexity}
\label{sec:magic_complexity}
In this section, we take the limit of $Z$-basis only measurements and examine the single-pair all-to-all model. Since $T$ and CZ are both diagonal in the $Z$-basis, $[T,\mathrm{CZ}]=0$. It can also be shown that
\begin{equation}
 T^{(j)}\frac{\Pi^{(j)}_{z,\pm}\ket{\psi}}{||\Pi^{(j)}_{z,\pm}\ket{\psi}||} = e^{i \phi_{\pm}}\frac{\Pi^{(j)}_{z,\pm}\ket{\psi}}{||\Pi^{(j)}_{z,\pm}\ket{\psi}||},
\label{eq:T-scaling}
\end{equation}
where $\phi_{+} = 0$ and $\phi_{-} = \pi/2$ and we can  ignore the global phase factor.
Since $T$ and $CZ$ commute, and there are no other gates that mix $Z$-basis states, a $T$-gate on qubit $i$ can be commuted forward in time until it meets a $Z$-measurement on the same qubit, at which point its effect is absorbed into an irrelevant global phase. Suppose a $T$-gate acts on qubit $i$ at time $t$, and the first $Z$-basis measurement after the $T$-gate occurs on qubit $i$ at time $t'>t.$ Then, if the last $Z$-basis measurement on qubit $i$ occurs at time $t'$, then all the $T$-gates on qubit $i$ at $t \leq t'$ are removed from the circuit.

To test the accuracy of the algorithm, with a truncation error threshold $\epsilon$, we evolve the LRSD under the circuit, truncated at each timestep to keep only the coefficients $\lambda_{l}$ such that $|\lambda_{l}| > \epsilon$. If there are no remaining logicals with $|\lambda_{l}| > \epsilon$, the trajectory is invalid and not considered. We pick $\epsilon$ sufficiently small so that this issue does not arise.

\begin{figure}[htbp]
    \centering
    \includegraphics[width=3.4in]{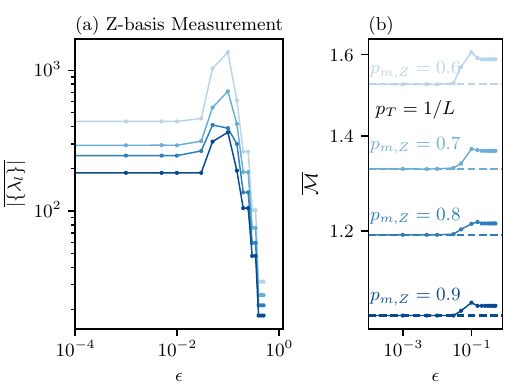}
    \caption{
        (a) $\overline{\left|\{\lambda_l\}\right|}$ as a function of truncation error $\epsilon$ for $\epsilon$ in $[10^{-4},0.5]$.  The $\overline{\cdots}$ denotes averaging over circuit trajectories. $|\{\lambda_{l}\}|$ represents the total number of logicals $\lambda_{l}$, where $\lambda_{l}$ is defined in Eq.~\eqref{eq:LSRD_decomposition}. Data is shown for the single-pair all-to-all model with $Z$-basis measurements, where Bell Sampling is performed on $\ket{\psi} \otimes \ket{\psi}$ for system size $L=50$, probability of $T$-gates $p_{T}=1/L$, and measurement probabilities $p_{m,Z} \in \{0.6, 0.7, 0.8, 0.9\}$, with a darker color denoting a larger measurement probability. (b)
        Stabilizer nullity $\overline{\mathcal{M}}$, as defined in Eq.~\eqref{eq:magic_equation_main}, as a function of the cutoff $\epsilon$ for several measurement probabilities $p_{m,Z}$. The dashed line denotes data obtained at zero truncation error. Data was obtained by averaging over $\sim 10^{3}$} circuit realizations. For each set of $\epsilon$ at a fixed $p_{m,Z}$, the locations of unitary gates and measurements are fixed. 
    \label{fig:sampling_efficiency}
\end{figure}

In Fig.~\ref{fig:sampling_efficiency}, we plot the dependence of the total number of logicals, $\overline{|\{\lambda_l\}|}$ on $\epsilon$ for varying $p_{m,Z}$. The $\overline{\cdots}$ denotes an average over circuit trajectories and $|\{\cdots\}|$ is the set size. We fix the location of the unitary gates and measurement locations to ensure $\overline{\left|\{\lambda_l\}\right|}(\epsilon)$ depends only on the truncation error and not the randomness of the circuit-to-circuit fluctuations. As shown in  Fig.~\ref{fig:sampling_efficiency}, as the cutoff is varied from $\epsilon = 0$ to $\epsilon \approx 10^{-2}$, the number of logicals is unchanged, which demonstrates that for this given circuit, this is the required number of logicals. In contrast, further increasing the cutoff $\epsilon$ to approach 1, $|\{\lambda_{l}\}|$ is reduced by roughly two orders of magnitude, or a factor of 100, in the effective Hilbert space dimension, shown in Fig.~\ref{fig:sampling_efficiency}(a). This greatly reduces the computational cost of Bell sampling at the price of a less accurate wavefunction.  We also clearly see the role of $p_{m,Z}$ measurements, which reduces the number of required logicals (a) and we find that (b) the stabilizer nullity grows by no more than five percent across the entire measurement rate window $p_{m,Z} \in \{0.6,0.7,0.8,0.9\}$ as compared to the zero truncation curves (dashed lines). These observations demonstrate that non-trivial observables remain robust to moderate truncation of the LRSD. This allows us to introduce  $\epsilon$ as an effective cutoff scheme to reduce simulation time while preserving quantitative accuracy. We note that when operating within the area-law magic phase where the number of logicals scales polynomially, $\epsilon$ can be safely set to machine precision for exact simulation, as was done for our space complexity analysis.

\begin{figure*}[htbp]
    \centering
    \includegraphics[width=7in]{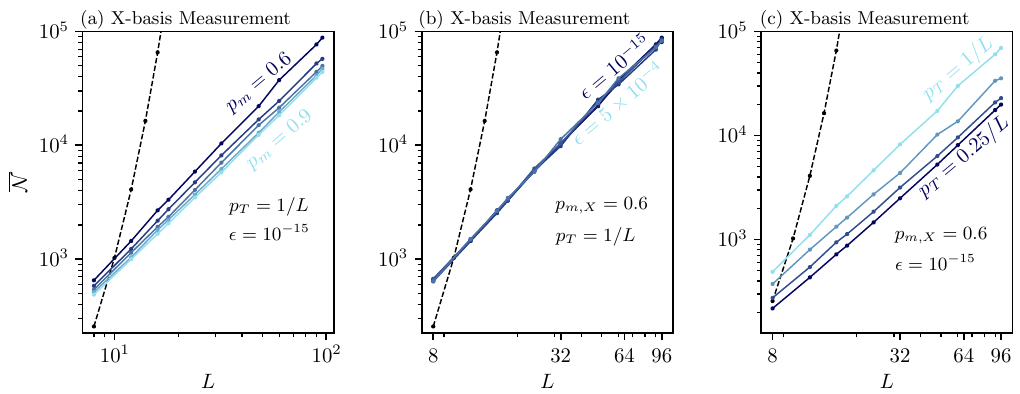}
    \caption{
        (a) Scaling of the circuit-averaged number of entries $\overline{\mathcal{N}}$ with $L$, see Eq.~\eqref{eq:NumberEntries}, which is the total number of entries required to specify the state in our algorithm. $|\{\lambda_{l}\}|$ represents the total number of logicals $\lambda_{l}$, where $\lambda_{l}$ is defined in Eq.~\eqref{eq:LSRD_decomposition}. Data is shown for the single-pair all-to-all model with $X$-basis measurements. For the state-vector simulation, $\mathcal{\overline{N}} = 2^{L}$, as shown by the black dashed line. Data is shown for measurement probabilities $p_{m,X} \in \{0.6,0.7, 0.8, 0.9\}$
        (b) Data is shown for cutoff values $\epsilon \in \{1 \times 10^{-15}, 1 \times 10^{-10},1 \times 10^{-6}, 1 \times 10^{-5}, 5 \times 10^{-4} \}$, with a darker color denoting a larger cutoff. (c) Data is shown for $T$-gate probabilities $p_{T} \in \{1/L,0.75/L,0.5/L,0.25/L \}$. Data was obtained by averaging over $\sim 10^{3}$} circuit realizations.
    \label{fig:log_scale}
\end{figure*}

Following the Algorithm in Sec \ref{sec:non-clifford-evolve} for applying $T$-gates, each $T$-gate will at most triple $\overline{\left|\{\lambda_l\}\right|}.$ This means, in the worst case scenario, $\overline{\left|\{\lambda_l\}\right|}=3^{|T|}$, where $|T|$ is the number of $T$-gates in the circuit. In this section, we consider a model where a single layer of $T$-gates is applied at the final timestep of the Clifford circuit. That is, we sweep from $i=1$ to $i=L$ once on a Clifford state. This means $\overline{\left|\{\lambda_l\}\right|}$ could, at worst, grow to $3^{L}$. The origin of this $3^L$ scaling comes from the fact that each of the $L$ $T$-gates can, in the worst case, triple the number of terms in the LRSD.
The rapid growth in $\overline{\left|\{\lambda_l\}\right|}$ imposes a challenge in the classical simulation. Therefore, only the state within the area law magic phase can be efficiently simulated using LRSD. In the area law magic phase, we find that the number of logicals, when averaged over circuits, scales with system size as a power law, as shown in Fig.~\ref{fig:log_scale}.

\subsection{Space Complexity}
\label{sec:space_complexity}
We now turn to the space complexity in our new simulation algorithm. To quantify the space complexity, we use the metric defined in Eq.~\eqref{eq:NumberEntries}. To emphasize the classical difficulty of simulating these circuits using the LRSD algorithm versus brute force state-vector methods, we now focus on the circuit-averaged number of entries $\overline{\mathcal{N}}$. The model we examine is the single-pair all-to-all model with $X$-basis measurements, as shown in Fig.~\ref{fig:circuit-schematic}. To demonstrate robustness to sufficiently generic circuits, we test our LRSD algorithm in Appendix~\ref{sec:space_appendix} using completely random Clifford gates in place of CZ gates.

We present the average number of entries $\overline{\mathcal{N}}$ as a function of $L$ in Fig.~\ref{fig:log_scale}(a), for $p_{m,X}$ varied from $0.6$ to $0.9$ in increments of $0.1$ at fixed $p_T = 1/L$. This parameter regime lies within the area-law magic phase shown in Fig.~\ref{fig:magic_transition_plot}. Due to the exponential growth of the logical space with the number of $T$-gates, the numerical simulation is constrained to approximately 100 to 200 $T$-gates at low measurement rates: at $p_T = 1/L$ the total $T$-gate count is $2L^2 p_T = 2L$, placing the system sizes shown in Fig.~\ref{fig:log_scale} within this window. We provide a comparison of $\overline{\mathcal{N}}=2^{L}$ state-vector methods (dashed black line) with the LRSD (solid blue lines). We find that $\overline{\mathcal{N}}$ exhibits power law growth with some exponent $\alpha$, $\overline{\mathcal{N}} = \beta + L^{\alpha}$, with $\alpha \approx 2$. This is much slower than the exponential growth of state-vector methods.
In the CAMPS representation, the computational cost remains polynomial for Clifford circuits and can often remain efficient when a Clifford circuit is “doped” with up to $\mathcal{O}(L)$ $T$-gates, particularly when these are distributed favorably across the system. In this regime, the disentangling procedure can frequently absorb the non-Clifford action without a significant increase in bond dimension \cite{Liu2025Classical}. Beyond this point—when the number of $T$-gates exceeds $\sim L$ or their placement generates unavoidable entanglement—the bond dimension typically grows rapidly, leading to exponential cost. For sufficiently sparse $T$-gate densities, such as rates scaling as $p_{T}$ $\sim 1/L^{2}$, the simulations remain tractable, and similar behavior is expected for small prefactors $\eta$ in the scaling form $p_{T}$ $\sim \eta/L$.
Within the area-law magic phase, the LRSD algorithm exhibits robust power-law scaling with system size $L$ (see Fig.~\ref{fig:log_scale}), which plausibly provides an upper bound on the scaling with $T$-gate count at fixed system size.

{We place our results in the context of prior work on classical simulation of Clifford circuits with a $T$-gate count $t$ proportional to the system size $L$. Prior works have demonstrated that circuits with $t \sim L$ $T$-gates can admit efficient simulation. It was shown \cite{Pashayan2022Fast} that the cost of exactly computing a chosen output probability scales as $\mathcal{O}(2^{t-r} \cdot t)$, where $r$ is a circuit-dependent quantity, termed the projector rank, that quantifies the number of linearly independent parity constraints induced by the $T$-gates. For random circuits with many Clifford gates between each $T$-gate, $r$ concentrates near its maximum $\min(t, n-w)$ (where $n$ is the number of qubits and $w$ the number of measured qubits), so the effective exponent $t-r$ remains small, enabling efficient simulation. Recent work \cite{Nakhl2025Stabilizer} has introduced a stabilizer tensor network method with magic state injection augmented stabilizer tensor network(MAST) and reported that local expectation values of random $T$-doped Clifford circuits can be computed in $\mathrm{Poly}(L)$ time when $t \lesssim L$, with a sharp phase transition in simulation cost beyond this regime. We note, however, that the $\mathrm{Poly}(L)$ cost in Ref.~\cite{Nakhl2025Stabilizer} applies to local expectation values; efficient sampling from the output distribution is explicitly not claimed.}

In light of these results, we emphasize that our contribution is not the observation that certain $t=\mathcal{O}(L)$ regimes can be tractable, but rather the identification and characterization of an average-case polynomial scaling regime for trajectory-resolved \emph{full density matrix} evolution in measurement-driven ensembles. Unlike the approaches above, which target output probabilities \cite{Pashayan2022Fast} or local expectation values \cite{Nakhl2025Stabilizer}, the LRSD maintains the complete density matrix $\rho(t)$, enabling the computation of nonlinear observables such as entanglement entropies and stabilizer nullity. This represents a fundamentally different and more demanding simulation objective.

To rigorously justify the average-case polynomial efficiency of the LRSD algorithm, it is critical to address the potential for a ``hard trajectory'' problem. This scenario is common in near-Clifford simulators where a small fraction of quantum trajectories incur exponential simulation costs, thereby dominating the ensemble average. As defined in Eq.~\ref{eq:NumberEntries}, the space complexity is determined by the number of logical operators $|\{\lambda_{\ell}\}|$. While the worst case cost for a single trajectory is theoretically bounded by $3^{|T|}$, our numerics confirm that the ensemble is not bottlenecked by rare, exponentially costly instances. As detailed in Appendix~\ref{sec:space_appendix}, the trajectory-resolved probability distribution of $\mathcal{N}$ is tightly concentrated around its mean, with a rapidly decaying right-hand tail. This rapid supression of high-cost trajectories ensures that the polynomial scaling of the ensemble-averaged complexity $\mathcal{N}$ in Fig.~\ref{fig:log_scale} is accurately representative of a typical circuit realization. Consequently, the simulation remains efficient in the area-law magic regime without being derailed by statistical outliers.

We note that the $X$-basis model---initialization in $\ket{+}^{\otimes L}$, evolution under diagonal unitaries (CZ and $T$), and measurement in the $X$-basis---is structurally related to Instantaneous Quantum Polynomial-time (IQP) circuits~\cite{bremner2010classical}, for which classical sampling from the output distribution is believed to be intractable under standard complexity-theoretic assumptions. Since maintaining the full density matrix is a strictly more demanding computational task than sampling, worst-case hardness of the LRSD simulation in this circuit class is expected. This complexity-theoretic expectation is consistent with the $3^{|T|}$ worst-case bound discussed above. Crucially, the mid-circuit projective measurements---absent in standard IQP circuits---provide the mechanism that collapses the LRSD and concentrates the trajectory-resolved cost around its polynomial mean, enabling average-case efficient simulation to coexist with worst-case hardness.

We summarize the efficient simulation regimes in Table~\ref{tab:efficiency_summary}. For $X$-basis measurements with CZ gates, $\overline{\mathcal{N}}$ scales as $\mathrm{poly}(L)$, as demonstrated numerically in Fig.~\ref{fig:log_scale}. To verify that this scaling is robust to circuit structure rather than an artifact of the CZ gate, we also test $X$-basis measurements with random Clifford gates replacing CZ (Fig.~\ref{fig:random_log_scaling} in Appendix~\ref{sec:space_appendix}), finding the same polynomial scaling. For the $Z$-basis model with CZ gates, $[T, \mathrm{CZ}] = 0$ and $T$-gates commute through $Z$-measurements (Sec.~\ref{sec:magic_complexity}), so the LRSD evolution is purely Clifford and $\overline{\mathcal{N}}$ reduces to the stabilizer tableau size $(2L+1)^{2}$; no numerical verification is needed.
\begin{table*}[bth]
\centering
\color{red}
\caption{Summary of efficient simulation regimes for $\overline{\mathcal{N}}$ from the LRSD algorithm.} 
\label{tab:efficiency_summary}
\begin{tabular}{cccccc}
\toprule
Measurement Basis & Gate Type & $T$-gate Rate & Measurement Rate & Scaling & Numerics \\
\midrule
$X$-basis & CZ & $p_{T} = \eta/L^{\beta}$, $\beta \geq 1$ & $p_{m} \gtrsim 0.6$ & $\mathrm{poly}(L)$ & Fig.~\ref{fig:log_scale} \\
$X$-basis & Random Clifford & $p_{T} = \eta/L^{\beta}$, $\beta \geq 1$ & $p_{m} \gtrsim 0.6$ & $\mathrm{poly}(L)$ & Fig.~\ref{fig:random_log_scaling} \\
$Z$-basis & CZ & any $p_{T}$ & any $p_{m}$ & $(2L+1)^{2}$ & -- \\
\bottomrule
\end{tabular}
\end{table*}
\begin{figure*}[htbp]
    \centering
    \includegraphics[width=7in]{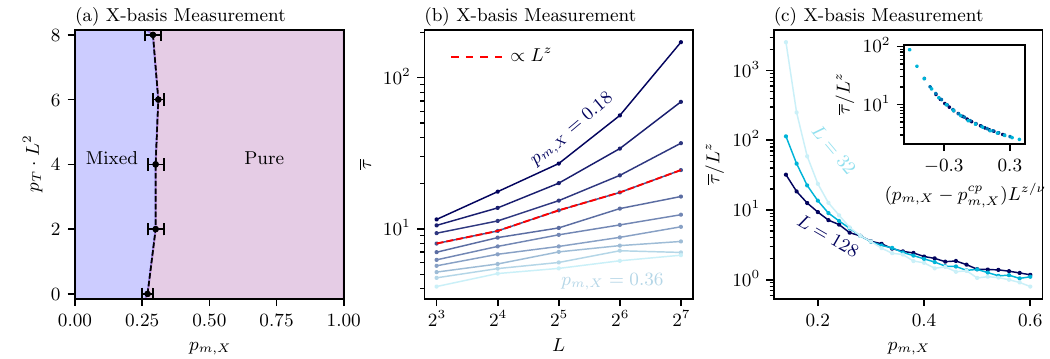}
    \caption{
(a) Phase diagram of the purification transition in the single-pair all-to-all model with $X$-basis only measurements as a function of measurement probability $p_{m,X}$ and probability of $T$-gates, $p_{T}$. The part of the phase diagram labeled mixed is the mixed phase of the purification transition, where $\overline{S_{Q}} \sim e^{t/\tau(L)}$ and $\overline{\tau(L)}$ grows exponentially in $L$. The part of the phase diagram labeled pure is the pure phase of the purification transition, where $\overline{\tau(L)}$ is the area law in $L$. $\overline{\cdots}$ denotes averaging over circuit trajectories.  The critical measurement probability separating the mixed phase and pure phase is given by the critical purification (cp) rate $p_{m,X}^{cp} = 0.26(1)$, which is weakly dependent on $p_{T}$. The points represent the numerically calculated locations of the purification transition, where $\tau(L)$ grows logarithmically in $L$. The black dashed line is the phase boundary deduced by interpolation.
(b) Purification time of $\overline{S_{Q}}$ as a function of system size $L$ at $p_T=0$ for different measurement probabilities $p_{m,X}$ from $0.18$ to $0.36$ in the steps of 0.02. The critical measurement probability is $p_{m,X}^{cp} = 0.26(1)$ with the dynamical exponent $z_{p}=0.22(2)$ (see Eq.~\eqref{eq:tau}). At $p_{m,X}^{cp}$, $\tau \sim L^{z_{p}}$, as indicated by the dashed line.
(c) $\tau L^{-z_p}$ as a function of measurement probability $p_{m,X}$ for $L\in \{32, 64, 128\}$. The inset shows the data collapse with the estimated critical exponent $\nu_p=0.45(5)$. The ensemble size is 25000 circuits for each value of $p_{m,X}$ and $L$ in all three panels.
}
\label{fig:all_to_all_purification_plot}
\end{figure*}

\section{Measurement-induced phase  transitions}
\label{sec:MIPT_transitions}

There are different patterns of multipartite entanglement in the all-to-all steady state, which can be distinguished by various metrics. In monitored circuits, three distinct entanglement phase transitions that can occur as the measurement rate $p_{m}$ is varied: (i) an entanglement transition at $p_{ce}$ which separates the volume- and area-law scaling, (ii) a connectivity (percolation) transition at $p_{cc}$ that fully disentangles the state, and (iii) a purification transition at $p_{cp}$ where a maximally mixed ancilla purifies in a time that becomes system size independent. Typically $p_{cp} \leq p_{ce} \leq p_{cc}$.

\subsection{Purification Transition}

Here, we focus on the purification transition in the single-pair all-to-all model with $X$-basis measurements and provide numerical evidence for its critical point and scaling. We defer detailed definitions and diagnostics for the connectivity and entanglement transitions to Appendix \ref{sec:EntanglementTransitions}. Typically $p_{cp} \leq p_{ce} \leq p_{cc}$. Numerically, we find that $p_{ce} = p_{cp} = 0.27(1)$ and $p_{ce} < p_{cc} = 0.66$.

The dynamical purification phase transition can be efficiently probed by studying how the system preserves entanglement over time with a single reference qubit. After preparing $\ket{\psi_{o}}=\ket{+x}^{\otimes L}$, the reference qubit is entangled to a randomly selected system qubit to form a Bell pair. The entangling operation is followed by a scrambling stage, which consists of $\sqrt{10}L$ random two-qubit Clifford gates applied to random system qubit pairs.
We choose $\sqrt{10}L$ gates because the system's all-to-all entanglement, Eq.~\eqref{eq:max-min_eq} in Appendix \ref{sec:entanglement_appendix}, is saturated at this point. This ensures that the ancilla is maximally entangled with all the qubits in the system, and not just a single qubit. This reduces the finite-size effects. We evolve the circuit up to a final time $t_f = 2L^2$ in the single-pair all-to-all model with $X$-basis measurements. This is shown in Fig.~\ref{fig:circuit-schematic}(a). This timescale is chosen to ensure that the ancilla entanglement entropy has entered its long-time regime, where it decays exponentially with time for all system sizes and measurement rates. The decay of $\overline{S}_{Q}$ with $t$, for $0<t<2L^{2}$ is shown in Fig.~\ref{fig:time_decay} at $p_{m,X}=p_{m,X}^{cp}$.

When we use $X$-basis measurements, Eq.~\eqref{eq:T-scaling} no longer applies. Recall that we used Eq.~\eqref{eq:T-scaling} in Sec.\ref{sec:magic_complexity} to argue that the $T$-gates constitute a single layer of single-qubit unitaries at the end of the Clifford circuit. The $T$ gates are commuted to the end of the circuit using the fact that they are diagonal in the $ Z$ basis and commute with both the $CZ$ gates and $Z$-basis measurements.
A final layer of single-qubit unitaries cannot alter the entanglement of a pure state $\ket{\psi}$. In this case, $\ket{\psi}$ is a stabilizer state, which has a degenerate entanglement spectrum. However,  since Eq. ~\eqref{eq:T-scaling} no longer applies, the argument at the beginning of Section \ref{sec:magic_complexity} is invalid. The $T$-gates are therefore interspersed throughout the circuit, which means the entanglement is no longer the same as the stabilizer state. Instead, the final state is a generic non-stabilizer state without a degenerate entanglement spectrum. We must therefore use the LRSD rather than the stabilizer formalism to simulate the state and calculate the entanglement.

We track the entanglement entropy between the system and the ancilla qubit, $\overline{S_{Q}(t)}$. $S_{Q}(t)$ is the von Neumann entanglement entropy, $S_{A}$, for a bipartition $A$ and $B$ of the system:
\begin{equation}
S_{A} = -\mathrm{tr}(\rho_{A} \mathrm{log} \rho_{A}).
\end{equation}
The purification time $\tau$ can be estimated from the decay of the ancilla entanglement, namely,  $\overline{S_{Q}(t)}\sim \exp(-t/\tau)$ (see Appendix \ref{sec:fitting_procedure} for details on the fitting procedure used to obtain $\tau$).
Therefore, by fitting the late-time decay and plotting $\tau$ as a function of system size, we can identify the purification transition, as shown in Fig.~\ref{fig:all_to_all_purification_plot}(b) and (c), leading to a phase diagram in Fig.~\ref{fig:all_to_all_purification_plot}(a) in the limit of $X$-basis only measurements, as a function of $p_{m,X}$ and $p_{T}$. Recall that $p_T = \eta/L^{\beta}$ in Eq.~\eqref{eq:magic_scaling} is the probability of injecting a single $T$-gate per timestep at a randomly chosen qubit (Sec.~\ref{sec:MIPT_transitions}); we choose $\eta=1$ and $\beta=2$
in Eq.~\eqref{eq:magic_scaling}. Each timestep contains $\eta/L^{\beta}$ $T$-gates on average, and accounting for $2L^{2}$ timesteps, this means that there are a total of $2L^{2}p_{T} = 2\eta L^{2-\beta}$
 $T$-gates in the circuit. For $L=64$, the number of $T$-gates is sufficiently low so that it fits in the constraints of our LRSD simulation method. In Section \ref{sec:clifford_algorithms}, we explain how to perform Clifford operations on the LRSD.

In the phase diagram Fig.~\ref{fig:all_to_all_purification_plot}, we find that the critical measurement rate $p_{m}^{cp}$ is unaffected by the rate of $T$-gates within error bars. This is consistent with prior work \cite{Fux2024Entanglement}, which has shown that the entanglement transition critical point $p_{c}^{ce}$ is insensitive to the rate of $T$-gates for $\beta \leq 1$, and we expect $p_{c}^{ce}=p_{c}^{cp}$ in this model.

In Fig.~\ref{fig:all_to_all_purification_plot}(b), we show the decay time as a function of system size.
At small measurement rates $p_m<0.25$, the initial state takes an exponentially long time to purify, indicating a mixed phase; whereas at large measurement rates $p_m>0.25$, the decay time saturates to an $ L$-independent value, indicating a pure state.
The critical measurement rate $p_{m,X}^{cp}= 0.26(1)$ can thus be identified by locating the inflection point of the decay time, where the decay time grows algebraically with system size, i.e.,
\begin{equation}\label{eq:tau}
    \tau \sim L^{z_p},
\end{equation}
where $z_p$ is the dynamical critical exponent for the purification transition.
We find $z_p = 0.22(2)$, which is consistent with the value found in Ref.~\cite{Skinner2019Measurement} for the Hartley entropy of Haar random circuits with measurements.
With the estimate of the critical measurement rate $p_{m,X}^{cp}$, and dynamical exponent $z_{p}$, we can also perform the finite size scaling of $\tau L^{-z_p}$ as a function of $p_m$ as shown in Fig.~\ref{fig:all_to_all_purification_plot}(c) using the following ansatz,
\begin{equation}\label{eq:scaling-ansatz}
    \tau L^{-z_p} \sim f_p((p_{m,X}-p_{m,X}^{cp})L^{z_{p}/\nu_p}),
\end{equation}
where $f_p(x)$ is a universal scaling function and $\nu_p$ is the critical exponent for the correlation length in the purification transition.
The estimated critical exponents are $\nu_p=0.45(5)$ and $z_p=0.22(2)$  with the data collapse as shown in the inset of Fig.~\ref{fig:all_to_all_purification_plot}(c).
Our estimate of the critical exponents is consistent with Ref.~\cite{Noel2022Observation} with long-range $XX$ gates, where $XX = \text{exp}(i\frac{\pi}{4}X\otimes X)$, implying the same universality class as mean-field percolation.

For the $n=0$ R\'enyi entropy of Haar random circuits with measurements, also known as the Hartley entropy, there is an exact mapping to a percolation problem in the circuit geometry. In an all-to-all geometry, this mapping predicts $z_{p}=1/5$ \cite{Nahum2021Measurement}.

Our primary focus is on the purification transition, which is about temporal correlations. For completeness, however, we also locate the entanglement transition, which is about spatial correlations. In the limit of no $T$-gates, we find in Appendix \ref{sec:ClusterTransition} that $p_{ce}=0.27(1)$, coinciding with the entanglement transition. We use an all-to-all definition of the entanglement suited for 0D systems.

The connectivity transition is analytically solvable for Haar random circuits in 1+1 dimensions, $p_{cc}=1/2$, due to the mapping onto 2D percolation \cite{Jian2020Measurement}. However, this mapping no longer accurately describes the connectivity properties of the state. Instead, the geometric structure of the entanglement correlations in the system is completely contained in the graph state $\ket{G}.$ The graph $G$ describes clusters of entangled qubits, which we use as the basis of our percolation analysis in Appendix \ref{sec:ClusterTransition}.

\subsection{Magic Transition}
\begin{figure*}[htbp]
    \centering
    \includegraphics[width=7in]{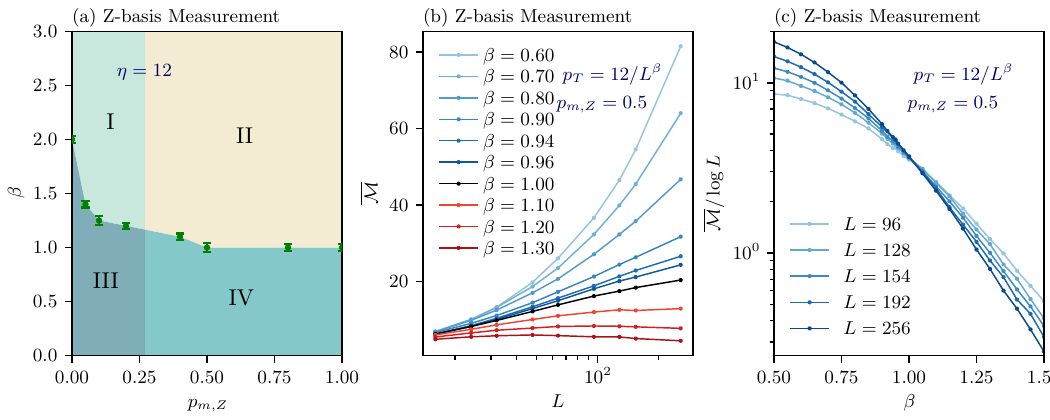}
    \caption{(a) Phase diagram of the magic and entanglement transitions in the single-pair all-to-all model with $Z$-basis only measurements as a function of the measurement probability $p_{m,Z}$ and $T$-gate rate $p_{T}=\eta/L^{\beta}$. The  numerical results are shown for $\eta=12$ and $p_{m,Z} \in \{0.0,\cdots,1.0\}.$ The green points show the numerically extracted locations of the magic transition. The entanglement transition is degenerate in the rate of $T$-gates, and was located using stabilizer simulations for $p_{T}=0.$ For low measurement rates and low $T$-gate rates, we find volume-law entanglement and area-law magic (phase I). For high measurement rates and low $T$-gate rates, we find area-law entanglement and area-law magic (phase II). For low measurement rates and high $T$-gate rates, we find volume-law entanglement and power-law magic (phase III). For high measurement rates and high $T$-gate rates, we find area-law entanglement and power-law magic (phase IV). (b) Scaling of the stabilizer nullity $\mathcal{\overline{M}}$, as defined in Eq.~\eqref{eq:magic_equation_main}, with $L$. The $\overline{\cdots}$ denotes averaging over circuit trajectories. The measurement probability is $p_{m,Z}=0.5$ and $0.6 \leq \beta \leq 1.3$ and $\eta=12$. Blue curves denote the (sub)-extensive phase, the black line denotes logarithmic criticality, and the red lines denote the area law phase. Data is shown in the single-pair all-to-all circuit model with $Z$-basis measurements. (c) $\mathcal{\overline{M}}$ divided by $\mathrm{log}(L)$ as a function of $\beta$ for different system sizes.
        }
    \label{fig:magic_transition_plot}
\end{figure*}
Here, we analyze a magic transition in the single-pair all-to-all model with only $Z$-basis measurements. Following the argument around Eq.~\eqref{eq:T-scaling}, $Z$-basis measurements constrain the placement of $T$-gates in a structured way. This leads to a density matrix representation that is amenable to Bell sampling, as discussed in Appendix \ref{sec:BellSample}. We then apply the algorithms developed in Appendix \ref{sec:BellSample} to efficiently compute the magic.
The phase transition is between a (sub)-extensive and area law scaling of magic controlled by the rate of measurements $p_{m,Z}$ and probability of $T$-gates $p_{T}$.  To find a phase transition in the magic, we study states that interpolate between states produced by Clifford and non-Clifford circuits in a controlled way.
We set the probability to sample $T$ gates as
\begin{equation}
p_{T} = \eta/L^{\beta}
\label{eq:magic_scaling}
\end{equation}
Previous work in a bricklayer geometry has shown that if the density of non‑Clifford $T$ gates is kept constant as the system size $L$ grows, $\beta = 0$, the total magic follows a volume law for any measurement rate $p < 1$ \cite{Fux2024Entanglement}. Magic is additive and can be produced locally, whereas entanglement can only be generated by gates whose support straddles the boundary between two subsystems. Because each site receives a fixed injection of magic from a finite gate density, the total magic therefore scales extensively with $L$ for $\beta = 0$. For $\beta=1$, $\eta$ represents average number of $T$ gates per time step, independent of $L$. The resulting density is $p_{T} = \eta / L^{\beta}$; choosing $\beta = 1$ drives $p_{T} \to 0$ in the thermodynamic limit while still providing a finite reservoir of magic.

When we add $T$-gates to the circuit in this model, the entanglement entropy does not change for each \textit{circuit}, because the $T$-gates constitute a single layer of single-qubit unitaries at the end of the circuit. Any sequence of single-site unitaries acting on an initial state cannot alter the entanglement across any bipartition; the entanglement in the final state is identical to that of the initial state. This means that the entanglement entropy of the circuit with $T$-gates can be calculated by simulating the stabilizer circuit with the $T$-gates removed. It also means that the magic phase transition cannot be detected by the entanglement, since it is degenerate in the number of $T$-gates. We emphasize that this follows from the fact that the $T$-gates, CZ gates, and $Z$-basis measurements are all diagonal in the computational basis and therefore mutually commute. As a result, the effective action of the $T$-gates can be represented as a single layer applied at the final stage of the circuit.

The observed magic transition depends solely on the locations of measurements and $T$-gates, which are chosen stochastically. For each circuit realization, the amount of magic is determined by the measurement history, which dictates how many $T$-gates survive until the end of the circuit. The final state along a given trajectory is a stabilizer state dressed by a measurement-conditioned static layer of $T$-gates.

The entanglement dynamics in this model are governed entirely by the underlying monitored Clifford circuit. The $T$-gates do not modify the scrambling-measurement competition between the Clifford entangling operations and measurements. Conversely, the influence of $T$-gates on magic is completely independent of the Clifford unitaries. Magic depends exclusively on which $T$-gates survive.

Across many trajectories, the survival probability of $T$-gates depends on the measurement rate, giving rise to a phase transition. There is no dynamical interplay between entanglement growth and the propagation of non-Clifford resources. Instead, the distribution of surviving non-Clifford resources, and thus the scaling of the stabilizer nullity, remains a dynamical property of the monitored circuit ensemble. Although $T$-gates commute forward through the circuit, their survival probability depends on the stochastic and dynamical measurement history.

We locate the magic transition using two methods. First, we fit the data to the function
\begin{equation}
f(L) = a + bL^{\gamma},
\end{equation}
with the constraint that $\gamma > 0$ corresponds to (sub)-extensive scaling, and $\gamma \leq 0$ to area-law behavior. If the best fit yields $\gamma \leq 0$, we classify the data as area-law; otherwise, we interpret it as (sub)-extensive. Details of the fitting procedure are provided in Appendix \ref{sec:magic_analysis}. The values of magic for the trajectories plotted in Fig.~\ref{fig:magic_transition_plot} empirically indicate a (sub)-extensive/log/area-law scaling of the magic with $L$. This is consistent with the scaling behavior of the magic, which was previously studied in bricklayer random Clifford circuits with a vanishing density of $T$-gates \cite{Fux2024Entanglement}. At the critical point $\beta = \beta_c$, we therefore invoke logarithmic scaling, i.e., $\mathcal{M}(L) = a + b \log L$ or
\begin{equation}
\frac{\mathcal{M}(L)}{\mathrm{log}(L)} = \frac{a}{\mathrm{log}(L)} + b.
\end{equation}
For $L \rightarrow \infty$, $\mathcal{M}/\mathrm{log(L)} \rightarrow b$, so we expect a crossing at $b$ in the thermodynamic limit.

In Fig.~\ref{fig:magic_transition_plot}(a), we identify four distinct phases in the magic and entanglement. The phases are area-law magic, volume-law entanglement (Phase I), area-law magic, area-law entanglement (Phase II), power-law magic, volume-law entanglement (Phase III), and area-law entanglement, power-law magic (Phase IV). In this model, the entanglement and purification transitions coincide. The exact exponent $\beta$ of the scaling plays a crucial role. Consider parameters $p_m$ and $\eta$ in regions III or IV. For some finite $\beta$, these points lie in the volume-law magic phase, while for $\beta = \infty$, no magic is present. We show numerically in Fig.~\ref{fig:magic_transition_plot}(a), that for sufficiently low $p_m$, $\beta_c > 1$ and decreases to $\beta_c=1$ for increasing $p_{m}$. This suggests that the $\beta = \infty$ ``zero-magic'' fixed point is stable. Conversely, for $p_m$ and $\eta$ in regions I or II, the system is in the area-law magic phase for some finite $\beta$, and is in the volume-law magic phase at $\beta = 0$. For sufficiently high $p_m$ and low $\eta$, it is likely that a critical point $\beta_c < 1$ exists.

We leave to future work as to whether, as $p_m \to 0$ and $\eta \to \infty$, the critical value $\beta_c$ also diverges. In other words, whether the $\beta = \infty$ ``zero-magic'' fixed point is unstable, implying that the system resides in the magic phase for any positive $\beta$. Similarly, it remains an open question whether, for sufficiently large $p_m$ and small $\eta$,  $\beta$ tends to zero, placing the system in the area-law magic phase for any finite $\eta$.

\section{Discussion and conclusion}

In this work, we introduced a simulation method based on the LRSD that stores and evolves the full density matrix of a near-stabilizer state. This capability allows access to quantities nonlinear in the state, such as entanglement. Using this approach, we identified, in the limit of $Z$-basis-only measurements, a MIPT in magic that is distinct from the MIPT in entanglement.

To compute stabilizer nullity for large systems ($L=256$), we developed a Bell sampling method. From a resource-theoretic perspective, we analyzed when LRSD offers a computational advantage, with the number of entries $\mathcal{N}$ providing a practical bound on simulation hardness as the $T$-gate rate varies. Introducing $T$-gates alters entanglement depending on the initial state and circuit structure. In $Z$-basis measurements within our CZ-based circuits, they can be commuted to the end and leave entanglement unchanged. With $X$-basis measurements, they can actively modify entanglement.

For the $Z$-basis measurement model, we find four distinct phases in entanglement and magic. Phase III begins generating sub-extensive entanglement and magic—resources essential for universal quantum computation—while Phases II and IV remain low-entanglement and can be efficiently simulated with MPS.
In the area-law magic regime (Phases I and II), the LRSD can further explore entanglement MIPTs in various circuit geometries and probe dynamical features such as operator growth and information scrambling via out-of-time-order correlators.

A limitation of our approach is that efficient simulation of the near-stabilizer $\rho$ for large ($L > 32$) system sizes is possible either near the critical measurement rate, which separates the area and (sub)-extensive magic phases, or within the area law magic phase. This constrains the values of $\beta$ and $\eta$ in the $T$-gate rate $p_{T}=\eta/L^{\beta}.$

Previous work has used MPS to study the magic transition in 1+1 dimensions using Clifford unitaries doped with $T$-gates. However, MPS fails for the nonlocal gates in the all-to-all model. Our simulator allows us to characterize the magic transition in this geometry.
Previous work has computed the entanglement transition in Haar-random all-to-all circuits \cite{Nahum2021Measurement} and the purification transition in Clifford all-to-all circuits \cite{Noel2022Observation}. Both of these works find similar scaling behavior, suggesting that the two universality classes are the same. Our results find critical exponents, $\nu = 0.5$ and $z=1/5$, as these work for various rates of $T$-gates, and confirm that the entanglement and purification transitions coincide in this all-to-all model.

In summary, our work presents new methodologies for simulating quantum circuits near the stabilizer limit and characterizing distinct MIPTs. To study the magic transition, we implemented a scalable Bell sampling algorithm capable of computing stabilizer nullity for large system sizes. The LRSD allowed us to study the purification in these circuits, which is independent of the magic transition. In other geometries, the purification transition may be nontrivially affected by the rate of $T$-gates and be related to the magic transition. Future research directions include extending this framework to non-Clifford resources beyond $T$-gates, and exploring MIPTs in other circuit geometries, where the universality class may undergo nontrivial flows.

\section*{Data Availability} The data used in this work is available from the authors upon request.

\acknowledgements{We thank David Huse, Grace Sommers, and Bryan Clark for useful discussions. This work was partially supported by the Army Research Office Grant No.~W911NF-23-1-0144 (K.A.~ and J.H.P.), the US-ONR grant No.~N00014-23-1-2357 (H.P. and J.H.P.).  K.A. and M.J.G. acknowledge support from  the Defense Advanced Research Projects
Agency (DARPA) under Agreement HR00112490357 and the NSF QLCI award OMA2120757. We are grateful to the OSG Consortium \cite{Sfiligoi2009Pilot,PordesOpen2007,OSG,OSPool} for providing us computational resources to run simulations. This work was  performed in part at the Aspen Center for Physics, which is supported by National Science Foundation grant PHY-2210452 (J.H.P.) and at the Kavli Institute for Theoretical Physics (KITP), which is supported by grant NSF PHY-2309135 (M.J.G., J.H.P.)}

\clearpage
\bibliography{References.bib}
\appendix

\section{Algorithms} \label{sec:algorithms}
In this section, we provide an algorithm to update the LRSD in Eq.~\eqref{eq:LSRD_decomposition} under Clifford operations. These are the Clifford unitaries and Pauli measurements. We also explain how to compute the Born probabilities of measurement outcomes, and update the state after the measurement. We also explain how to compute the R\'enyi entanglement entropy from the LRSD. In the final part, we develop algorithms to perform Bell Sampling on a specific class of non-Clifford states. We also present the method of computing the stabilizer nullity.
\subsection{Clifford Evolution} \label{sec:clifford_algorithms}
In this section, we describe how to update the LRSD of a quantum state as it evolves under Clifford unitaries and Pauli measurements, see Eqs.~\eqref{eq:CZ-gate} and ~\eqref{eq:clifford_measurement}.  Throughout, we write the density matrix in the form of Eq.~\eqref{eq:LSRD_decomposition} of the main text,
\begin{equation}
\ketbra{\psi}= \sum_{l\in\mathcal L_{\mathrm{LRSD}}} \lambda_{l}\sigma_{l}\rho_{S},
\end{equation}
where the coefficients $\lambda_{l}$ are real, the operators $\sigma_{l}$ are logical Pauli strings drawn from the set $\mathcal L_{\mathrm{LRSD}}$,
and $\rho_{S}$ is a stabilizer density matrix.  Clifford unitaries and Pauli measurements are resource‑free in the sense that they do not generate additional non‑Clifford resources, such as magic. Nevertheless, they alter both the logical operators and the stabilizer structure, so the LRSD must be updated after every such operation.

Let $\mathcal S=\expval{g_{1},\dots ,g_{r}}$ denote the abelian stabilizer group of $\rho_{S}$, generated by the commuting Paulis $g_{i}$.  For a mixed stabilizer state, $r < L.$ We then define a corresponding set of destabilizers $\mathcal D=\{d_{1},\dots ,d_{r}\}$ with the properties $[d_{i},d_{j}]=0$ and $[g_{i},d_{j}]=0$ for $i\neq j$, while each pair $(g_{i},d_{i})$ anticommutes, $\{g_{i},d_{i}\}=0$.  The set $\mathcal D$ complements the stabilizer generators but, unlike $\mathcal S$, does not form a group.
\subsubsection{Clifford Unitaries}
Equation~\eqref{eq:unitary_near_clifford} gives the update rule for a Clifford unitary $U_{C}$.  Because any Clifford maps Pauli operators onto Pauli operators by conjugation, each logical operator transforms as $\sigma_{l}\mapsto U_{C}\sigma_{l}U^{\dagger}$, and the stabilizer state transforms as $\rho_{S}\mapsto U_{C}\rho_{S}U_{C}^{\dagger}$.  Consequently the LRSD retains its form after the update: the set of stabilizer generators $\{g_{i}\}$, the destabilizers $\{d_{i}\}$, and the logical Paulis $\{\sigma_{l}\}$ are merely mapped onto other Pauli operators by the conjugation, while the coefficients $\lambda_{l}$ remain unchanged. The LRSD is then updated as:
\begin{equation}
\label{eq:unitary_near_clifford}
\ketbra{\psi}
\mapsto
\sum_{l \in \mathcal{L_{\mathrm{LRSD}}}} \lambda_{l}
U_{C} \sigma_{l} U_{C}^{\dagger}
U_{C} \rho_{S} U_{C}^{\dagger}.
\end{equation}
\subsubsection{Pauli Measurements}
We now discuss how to perform a projective measurement of the Pauli operator $P$, where
\begin{equation}
\label{eq:PauliOperator}
P = \bigotimes_{j=1}^{L} \sigma_{ij}, \qquad \sigma_{ij}\in\{I, X, Y, Z\}.
\end{equation}
The effect of the measurement depends on how $P$ commutes with elements of $\mathcal{L}_{\mathrm{LRSD}}$ and $\mathcal{D}.$ We use the notation $[P,\mathcal{S}]=0$ to mean that $P$ commutes with all elements of $\mathcal{S}$, and $[P,\mathcal{S}]\neq0$ to mean that $P$ anticommutes with at least one element of $\mathcal{S}.$ We recall that $P$ either commutes or anticommutes with every other Pauli operator. The algorithm then branches into three cases. When $[P,\mathcal{L}_{\mathrm{LRSD}}]=0$, the stabilizer state is updated directly. When $[P,\mathcal{L}_{\mathrm{LRSD}}]\neq0$ but $[P,\mathcal{S}]=0$, certain logical operators are dropped. When $[P,\mathcal{L}_{\mathrm{LRSD}}]\neq0$ but $[P,\mathcal{S}]\neq0$, a stabilizer decomposition is performed before applying the measurement.

By a stabilizer decomposition, we mean that
\begin{equation}
\label{eq:Decomposition}
\rho_{S} = \frac{1+\bar{P}}{2}\rho_{S}^{P},
\end{equation}
where $\rho_{S}^{P}$ is a new stabilizer state such that $[\rho_{S}^{P},P] = 0$ and $\bar{P}$ is a Pauli operator defined analogously to Eq.~\eqref{eq:PauliOperator}.
\begin{algorithm}[H]
\caption{Pauli Measurements}
\begin{algorithmic}

    \State For a measurement of Pauli $P$ with the projector
    $\dfrac{1 \pm P}{2}$,
    the state will evolve as:
\end{algorithmic}
\[
\ketbra{\psi} \rightarrow \sum_{l \in \mathcal{L}_{\mathrm{LRSD}}} \lambda_{l} \frac{1 \pm P}{2} \sigma_{l} \rho_{S} \frac{1 \pm P}{2}.
\]
\begin{algorithmic}
    \State There are three cases:
    \State \textbf{Case I:} $[P, \mathcal{L}_{\mathrm{LRSD}}] = 0$
    \State In this case, update $\rho_{S}$:
\end{algorithmic}
\[
\rho_{S} \rightarrow \frac{1 \pm P}{2} \rho_{S} \frac{1 \pm P}{2}
\]
\begin{algorithmic}
    \State \textbf{Case II:} $[P, \mathcal{L}_{\mathrm{LRSD}}] \neq 0$, but $[P, \mathcal{S}] = 0$
    \State Update the state as follows:
\end{algorithmic}
\[
\ketbra{\psi} \rightarrow \sum_{l \in \mathcal{L}_{\mathrm{LRSD}}} \lambda_{l} \frac{1 \pm P}{2} \sigma_{l} \frac{1 \pm P}{2} \rho_{S}
\]
\begin{algorithmic}
    \State If $\{P, \sigma_{l}\} = 0$, drop that term from the sum:
\end{algorithmic}
\[
\ketbra{\psi} \rightarrow \sum_{\substack{l \in \mathcal{L}_{\mathrm{LRSD}} \\ [P, \sigma_{l}] = 0}} \lambda_{l} \sigma_{l} \frac{1 \pm P}{2} \rho_{S}
\]
\begin{algorithmic}
    \State Calculate the probability of the outcome $\pm P$ using:
\end{algorithmic}
\[
\text{tr}(P \ketbra{\psi})
\]
\begin{algorithmic}
    \State \textbf{Case III:} $[P, \mathcal{L}_{\mathrm{LRSD}}] \neq 0$ and $[P, \mathcal{S}] \neq 0$
    \State Perform a decomposition of $\rho_{S}$ using Algorithm \ref{alg:decomposition} to obtain $\bar{P}$ and $\rho_{S}^{P}$.
    \State The state evolves as:
\end{algorithmic}
\[
\rho \rightarrow \sum_{\substack{l \in \mathcal{L}_{\mathrm{LRSD}} \\ [P, \sigma_{l}] = 0}} \frac{\lambda_{l}}{2} \sigma_{l} \frac{1 \pm P}{2} \rho_{S}^{P} + \sum_{\substack{l \in \mathcal{L}_{\mathrm{LRSD}} \\ [P, \sigma_{l}] \neq 0}} \frac{\lambda_{l}}{2} \sigma_{l} \bar{P} \frac{1 \pm P}{2} \rho_{S}^{P}
\]
\label{alg:msm_near_clifford}
\end{algorithm}

The proof of the evolution of $\rho$ in Case III is as follows. We give the proof for a projection onto $\frac{I+P}{2}$; the case for $\frac{I-P}{2}$ follows the same argument. First, using Algorithm  \ref{alg:decomposition}, we can decompose $\rho_{S} = \frac{1 +\bar{P}}{2} \rho_{S}^{P}$, where $[\rho_{S}^{P},P] = 0$ and $\{\bar{P},P\} = 0$. Then,
\begin{equation}
\label{eq:FULL_EXPANSION}
\begin{aligned}
\rho &= \frac{I + P}{2} \left( \sum_{l \in \mathcal{L}_{\mathrm{LRSD}}} \lambda_{l} \sigma_{l} \rho \right) \frac{I + P}{2} \\
     &= \frac{I + P}{2} \left( \sum_{l \in \mathcal{L}_{\mathrm{LRSD}}} \lambda_{l} \sigma_{l} \frac{I + \bar{P}}{2} \rho_{S}^{P} \right) \frac{I + P}{2} \\
     &= \sum_{l \in \mathcal{L}_{\mathrm{LRSD}}} \lambda_{l} \frac{I + P}{2} \sigma_{l} \frac{I + \bar{P}}{2} \frac{I + P}{2} \rho_{S}^{P}
\end{aligned}
\end{equation}
There are two cases for each term labeled by $l$ in the sum. If $[P, \sigma_{l}] = 0$, then we replace that term by
\begin{equation}
\label{eq:CASEI_expansion}
\begin{aligned}
\lambda_{l} \sigma_{l} \frac{I + P}{2} \frac{I + \bar{P}}{2} \frac{I + P}{2} \rho_{S}^{P}
&= \frac{\lambda_{l}}{2} \sigma_{l} \frac{I + P}{2} \rho_{S}^{P}
\end{aligned}
\end{equation}
Note that $\frac{I+P}{2} \rho_{S}^{P}$ defines a new stabilizer state $\rho_{S}.$
If $\{\sigma_{l},P\}=0$, then the term is
\begin{equation}
\label{eq:CASEII_expansion}
\begin{aligned}
\lambda_{l} \sigma_{l} \frac{I - P}{2} \frac{I + \bar{P}}{2} \frac{I + P}{2} \rho_{S}^{P}
&= \frac{\lambda_{l}}{2} \sigma_{l} \bar{P} \frac{I + P}{2} \rho_{S}^{P}
\end{aligned}
\end{equation}
Substituting Eqs.~\eqref{eq:CASEI_expansion} and ~\eqref{eq:CASEII_expansion} into each term of Eq.~\eqref{eq:FULL_EXPANSION}, we see that we arrive at the mapping in Case III.
To evaluate whether $\{P, \mathcal{L}_{\mathrm{LRSD}}\} = 0$ in Algorithm \ref{alg:update_algorithm}, and all other commutation relations, we use the symplectic inner product. A generic Pauli operator $P_{i}$ can be written as
\begin{equation}
P_{i} = \bigotimes_{j=1}^{L} X^{x_{ij}} Z^{z_{ij}},
\end{equation}
where $x_{ij}, z_{ij} \in \{0, 1\}$, where $i$ labels which Pauli operator it is.
The symplectic inner product for two Pauli operators $P_{1}$ and $P_{2}$ is defined as:
\begin{equation}
P_{1} \circ P_{2} = z_{11}x_{21} \oplus \cdots \oplus z_{1L}x_{2L} \oplus x_{11}z_{21} \oplus \cdots \oplus x_{1L}z_{2L},
\end{equation}
where $\oplus$ denotes addition modulo 2. This product equals 0 if $[P_{1}, P_{2}] = 0$ and 1 if $\{P_{1}, P_{2}\} = 0$.
We now present an approach that leverages the stabilizer structure to efficiently compute the Born probability of projective measurements. The algorithm computes the probability of obtaining the outcome $+1$ when performing a projective measurement of operator $P$ on the general pure state $\ket{\psi}.$ The routine projects the LRSD‐represented state $\ketbra{\psi}$ onto the $+1$ eigenspace of a Pauli $P$. First, each logical Pauli $\sigma_l$ that commutes with $P$ is rewritten as a difference of projectors, allowing the LRSD sum to be expressed entirely in terms of projector products. The trace of this projected operator yields the probability $P_{\uparrow,P}$ of obtaining outcome $+1$. $P_{\uparrow,P}$ can be interpreted as a sum of terms, where each term is a conditional probability of obtaining two consecutive measurements with fixed outcomes. Summing over the terms reconstructs the total measurement probability.

\begin{algorithm}[H]
\caption{Born Probabilities}
\begin{algorithmic}
    \State \textbf{Input:}
\end{algorithmic}
\[
\ketbra{\psi} \rightarrow \sum_{\substack{l \in \mathcal{L} \\ [P, \sigma_{l}] = 0}} \lambda_{l} \sigma_{l} \frac{1 \pm P}{2} \rho_{S}
\]
\begin{algorithmic}
    \State \textbf{Step 1:} Replace each $\sigma_{l}$ with the difference of projectors:
\end{algorithmic}
\[
\ketbra{\psi} \rightarrow \sum_{\substack{l \in \mathcal{L} \\ [P, \sigma_{l}] = 0}} \lambda_{l} \frac{1 + \sigma_{l}}{2} \frac{1 + P}{2} \rho_{S} - \sum_{\substack{l \in \mathcal{L} \\ [P, \sigma_{l}] = 0}} \lambda_{l} \frac{1 - \sigma_{l}}{2} \frac{1 + P}{2} \rho_{S}
\]
\begin{algorithmic}
    \State \textbf{Step 2:} Compute $P_{\uparrow, P}$ as the trace of this object:
\end{algorithmic}
\[
P_{\uparrow, P} = \text{tr}\left(\frac{1 + \sigma_{l}}{2} \frac{1 + P}{2} \ketbra{\psi}\right)
\]
\begin{algorithmic}
    \State \textbf{Step 3:} Define the state:
\end{algorithmic}
\[
\ket{\psi'} = \frac{\frac{1 + P}{2}\ket{\psi}}{\sqrt{P_{\uparrow,P}}}
\]
\begin{algorithmic}
    \State \textbf{Step 4:} Express the original trace as:
\end{algorithmic}
\[
\text{tr}\left(\frac{1 + \sigma_{l}}{2}\ketbra{\psi'}\right) P_{\uparrow, P} = P_{\uparrow, l} P_{\uparrow, P}
\]
\begin{algorithmic}
    \State \textbf{Step 5:} Therefore, compute $P_{\uparrow, P}$ as:
\end{algorithmic}
\[
P_{\uparrow, P} = \sum_{\substack{l \in \mathcal{L} \\ [P, \sigma_{l}] = 0}} \lambda_{l}(P_{\uparrow, l} - P_{\downarrow, l})P_{\uparrow, P}
\]
\end{algorithm}

We now present the Algorithm for performing a stabilizer decomposition as in Eq.~\eqref{eq:Decomposition}. Firstly, $\rho_{S}^{P}$ is obtained from $\rho_{S}$ by removing a stabilizer generator $\overline{P}$ satisfying $\{\bar{P},P\}=0.$ To see how this decomposition is obtained, we have
\begin{equation}
\rho_{S} = \frac{1}{2^{L-r}} \prod_{i=1}^{r} \frac{I+g_{i}}{2}.
\end{equation}
By a change of basis, we can choose at most one $g_{i}$ to anticommute with $P.$ The details of the procedure are given in Algorithm \ref{alg:decomposition}. Without loss of generality, suppose that $\{g_{1},P\}=0$ and $[g_{i},P]=0$ for all other $i.$ Then,
\begin{equation}
\rho_{S} = \frac{I + g_{1}}{2} \left[\frac{1}{2^{L-r}} \prod_{i=2}^{r} \frac{I + g_{i}}{2} \right].
\end{equation}
Identifying $g_{1} = \overline{P}$ and
\begin{equation}
\rho_{S}^{P} = \prod_{i=2}^{r} \frac{I+g_{i}}{2},
\end{equation}
we have Eq.~\eqref{eq:Decomposition}.

We now explain how the state $\rho_{S}^{P}$ is stored in stabilizer simulations and is obtained from the representation of the original state $\rho_{S}.$ For a mixed stabilizer state $\rho_S$, the stabilizer formalism includes not only stabilizers and destabilizers, but also a set of logical operators denoted by $\mathcal{L}_S = \{l^{x}_1, \dots, l^{x}_{L-r}, l_{1}^{z}, \cdots, l^{z}_{L-r}\}$. We group the logical operators into two blocks: the first $L-r$ operators denoted with $x$ and their anticommuting partners denoted with $z$. These logical operators satisfy the commutation relations $[l^{x}_i, l^{x}_j] = [l^{z}_i, l^{z}_j] = [l^{x}_i, l^{z}_j] = 0$ for $i \neq j$ and $\{l^{x}_i, l^{z}_i\} = 0$. Additionally, the logical operators commute with all stabilizers $s_j \in \mathcal{S}$ and destabilizers $d_j \in \mathcal{D}$, i.e., $[l^{x}_i, s_j] = [l^{z}_i, s_j] = [l^{x}_i, d_j] = [l^{z}_i, d_j] = 0$ for all $i, j$.

\begin{algorithm}[H]
\caption{Stabilizer Decomposition}
\label{alg:decomposition}
\begin{algorithmic}
    \State \textbf{Input:} Given stabilizers $g_{1},\dots,g_{r}$ with their corresponding destabilizers $d_{1},\dots,d_{r}$, logical operators $l^{x}_{1},\dots,l^{x}_{L-r}$ and $l^{z}_{1},\dots,l^{z}_{L-r}$, and a Pauli operator $P$, we assume—without loss of generality—that $g_{1}$ is the first stabilizer that anticommutes with $P$.

    \State \textbf{Step 1:} Replace $d_{1}$ with $g_{1}=\bar{P}$

    \State \textbf{Step 2:} For $j = 1, \cdots, r$, if $\{l_{j}^{x},P\}=0$, then $l_{j}^{x} \mapsto \bar{P} \cdot l_{j}^{x}.$ Similarly, if  $\{l_{j}^{z},P\}=0$, then $l_{j}^{z} \mapsto \bar{P} \cdot l_{j}^{z}.$
    \State \textbf{Step 3:} Remove $g_{1}$ and $d_{1}.$ Append $P$ to $\mathcal{L}$ so $l^{x}_{r+1}=P$ and append $\bar{P}$ to $\mathcal{L}$ so that $l^{z}_{r+1}=\bar{P}.$

    \State \textbf{Step 4:} Update $r$ to $r-1$.
\end{algorithmic}
\end{algorithm}

\subsection{Partial Trace and R\'enyi Entropy of LRSD}

We now present algorithms to compute the R\'enyi entanglement entropies using the representation of the state in Eq.~\eqref{eq:LSRD_decomposition}. We present an algorithm for computing the reduced density matrix for an arbitrary state, from which we can compute the entanglement. We also present an algorithm which is valid only for $[\mathcal{L}_{S},\mathcal{S}]=0.$ These methods allow us to compute the full entanglement spectrum for an arbitrary $n$, where:
\begin{equation}
\label{eq:entanglement_arb_n}
S_{n} = \frac{1}{1-n} \mathrm{log\:Tr}[\rho_{A}^{n}],
\end{equation}
and $\rho_{A}$ is the reduced density matrix of some region $A.$

Our method to compute the entanglement relies on converting the LRSD representation of $\rho$ in the compact binary form into a $2^{|A|} \times 2^{|A|}$ matrix in the full Hilbert space. This restricts the size of $|A|$ for which we can compute the entanglement. However, non-trivial quantities such as correlation functions between two qubits and the ancilla entanglement entropies can be computed.

As a concrete example, we demonstrate how to compute correlation functions between an arbitrary pair of qubits using the LRSD representation. Given any two qubits $A$ and $B$, we trace out all remaining qubits $k \notin \{A,B\}$ (using Algorithm~4) to obtain the reduced density matrix $\rho_{A \cup B} = \mathrm{Tr}_{k}(\rho)$, which is a $4 \times 4$ matrix. From this reduced density matrix, one can evaluate any two-qubit observable, including standard two-point correlators $\langle O_A O_B \rangle$. A particularly useful measure is the mutual information,
\begin{equation}
I_{n}(A \cup B) = S_{n}(\rho_{A}) + S_{n}(\rho_{B}) - S_{n}(\rho_{A \cup B}),
\label{eq:MutualInformation}
\end{equation}
which captures the total correlation---both quantum and classical---between any two arbitrary sites in the system. Here, $S_{n}$ denotes the R\'enyi entropy defined in Eq.~\ref{eq:entanglement_arb_n}. The mutual information provides an upper bound on all connected two-point correlation functions via $\langle O_A O_B \rangle - \langle O_A \rangle \langle O_B \rangle \leq \sqrt{2 I_1(A \cup B)}$, and thus serves as a comprehensive diagnostic of two-site correlations accessible within our framework.

\subsubsection{Partial Trace and R\'enyi Entanglement Algorithms}
We now present a general-purpose algorithm to compute the R\'enyi entanglement entropies for arbitrary $n$. Given region $A$ with qubits $\{i_{1},\cdots,i_{k}\}$ and region $B$ with qubits $\{i_{1},\cdots,i_{L-k}\},$ the first step is to perform a partial trace of the state in Eq.~\eqref{eq:LSRD_decomposition} to obtain the reduced density matrix
\begin{equation}
\rho_{A} = \sum_{\ell'} \lambda_{\ell'}^{A} \sigma_{\ell'}^{A} \rho_{S}^{A},
\end{equation}
where $\rho_{S}^{A}$ is the reduced density matrix of $\rho_{S}$ and $\sigma_{\ell'}^{A}$ are the logical operators restricted to region $A$. The size of the new logical set $\{\lambda_{\ell'}\}$ satisfies $|\{\lambda_{\ell'}\}| \leq |\{\lambda_{\ell}\}|.$
Each of the logical operators $\sigma_{\ell}$ and $\rho_{S}^{A}$ can then be converted to the full Hilbert space, represented as $2^{2k} \otimes 2^{2k}$ matrices, respectively. Using matrix operations, the matrix for the full state $\rho_{A}$ can be computed, from which Eq. ~\eqref{eq:entanglement_arb_n} can be easily evaluated.

The first step brings the tableau for stabilizer state $\rho_{S}$ into row-echelon form using the procedure of Ref.~\cite{audenaert2005entanglement}, which applies permutations and scalar multiplications to the generators. This form is particularly well-suited for partial trace operations and simplifies handling both pure and mixed stabilizer states.

We now present a method to perform a partial trace of $\rho$, which is represented using the LRSD in Eq.~\eqref{eq:LSRD_decomposition}. This method extends the technique in Ref. \cite{audenaert2005entanglement}, which shows how to compute the entanglement in stabilizer states. First, we represent the stabilizers of $\rho_{S}$ of Eq.~\eqref{eq:psi_decomposition} in an $r\times L$ matrix, where columns $1$ to $L$ represent qubits $1$ to $L$ and each row represents a different stabilizer.
Each entry of the matrix is either the Identity or Pauli operator: $I$,$X$,$Y$,$Z$. Each logical operator $\sigma_{l}$ is represented analogously. The first step is to bring the tableau of $\rho$ into a reduced row echelon form (RREF), following the procedure in \cite{audenaert2005entanglement}. The general structure of the RREF can be described in a recursive fashion. There are three cases:
\begingroup
  \setlength{\abovedisplayskip}{6pt}
  \setlength{\belowdisplayskip}{6pt}
  \setlength{\arraycolsep}{1pt}
  \renewcommand{\arraystretch}{1}
  \[
  \begin{array}{c@{\hspace{12pt}}c@{\hspace{12pt}}c}

    \begin{array}{c}
      \textbf{Case I}\\[2pt]
      \left(
        \begin{array}{@{\;}c@{\;}|@{\;}c@{\;}}
          1      &               \\
          \vdots & \mathrm{RREF}'\\
          1      &
        \end{array}
      \right)
    \end{array}

    &
    \begin{array}{c}
      \textbf{Case II}\\[2pt]
      \left(
        \begin{array}{@{\;}c@{\;}|@{\;}cccc@{\;}}
          \sigma & * & \cdots & * &        \\
          1      &   &        &   &        \\
          \vdots &   & \mathrm{RREF}' & &  \\
          1      &   &        &   &
        \end{array}
      \right)
    \end{array}

    &
    \begin{array}{c}
      \textbf{Case III}\\[2pt]
      \left(
        \begin{array}{@{\;}c@{\;}|@{\;}cccc@{\;}}
          \sigma_{1} & * & \cdots & * &    \\
          \sigma_{2} & * & \cdots & * &    \\
          1          &   &        &   &    \\
          \vdots     &   & \mathrm{RREF}' & &\\
          1          &   &        &   &
        \end{array}
      \right)
    \end{array}

  \end{array}
  \]
\endgroup
The symbols ``$\vdots$'' and ``$\cdots$'' denote a number of repeated rows and columns. This number can be zero. The symbol RREF denotes a possibly empty sub-array that is also in RREF form. The symbol $*$ denotes either a Pauli operator or $I$. Furthermore, $\sigma$, $\sigma_{1}$, and $\sigma_{2}$ are Pauli operators and $\sigma_{1}$ and $\sigma_{2}$ anticommute. We denote the sequences of $*$ operators by $g$, $g_{1}$ and $g_{2}$, respectively.
The first column in each of the three cases represents a qubit to be traced over. The stabilizer in RREF' defines a new stabilizer state $\rho'$ which is the reduced density matrix obtained after tracing over the first qubit. In each of the three cases, we can decompose $\rho$ into its action on the qubit to be traced over and the rest of the system. In Case I, recalling that we can substitute the generators $g_{i}$ from the tableau into Eq.~\eqref{eq:stabilizerstate}, we see that:
\begingroup
  \setlength{\abovedisplayskip}{6pt}
  \setlength{\belowdisplayskip}{6pt}
  \begin{equation}
  \label{eq:case1}
    \rho = \frac{1}{2} I \otimes \rho'.
  \end{equation}
\endgroup
In Case II, we can also use Eq.~\eqref{eq:stabilizerstate} to observe that
\begingroup
  \setlength{\abovedisplayskip}{6pt}
  \setlength{\belowdisplayskip}{6pt}
  \begin{equation}
  \label{eq:case2}
  \begin{aligned}
    \rho
      &= \left(\frac{I\otimes I + \sigma\otimes g}{2}\right) I\otimes \rho' \\
      &= \frac{1}{2}\bigl(I\otimes \rho' + \sigma\otimes g\rho'\bigr)
  \end{aligned}
  \end{equation}
\endgroup
Finally, in Case III, we have that
\begingroup
  \setlength{\abovedisplayskip}{6pt}
  \setlength{\belowdisplayskip}{6pt}
  \begin{equation}
  \label{eq:case3}
  \begin{aligned}
    \rho
      &= \left(\frac{I\otimes I + \sigma_{1}\otimes g_{1}}{2}\right) \left(\frac{I\otimes I + \sigma_{2}\otimes g_{2}}{2}\right)(I\otimes 2\rho') \\
      &= \frac{1}{2}\bigl(I\otimes \rho' + \sigma_{1}\otimes g_{1}\rho' + \sigma_{2}\otimes g_{2}\rho' + \sigma_{1}\sigma_{2}\otimes g_{1}g_{2}\rho'\bigr).
  \end{aligned}
  \end{equation}
\endgroup
We must have that $g_{1} \neq g_{2}$, since $g_{1} = g_{2}$ requires $\sigma_{1} = \sigma_{2}$ to preserve the property that all stabilizers commute. However, $g_{1} = g_{2}$ and $\sigma_{1} = \sigma_{2}$ leads to duplicate stabilizer generators.
For each logical $\sigma_{l}$, we can simply decompose it into
\begin{equation}
\sigma_{l} = \sigma_{A} \otimes \sigma_{B},
\label{eq:log_dec}
\end{equation}
where $\sigma_{r}$ is the action of $\sigma_{l}$ on the rest of the system. Now, we can present our algorithm to calculate the partial trace of the LRSD.
\\ \noindent\textbf{Step~1.} Using column permutations, bring the columns of the qubits to be traced out to the first position. Do this for each of the logicals~$\sigma_{l}$ and the tableau of~$\rho_{S}$.
\\ \noindent\textbf{Step~2.} Bring the stabilizer tableau into RREF$'$.
\\ \noindent\textbf{Step~3.} For each qubit to be traced over, apply Algorithm~\ref{alg:partial_trace_near_clifford}.

\begin{algorithm}[H]
\caption{Partial Trace}
\label{alg:partial_trace_near_clifford}
\begin{algorithmic}
    \normalsize

    \Statex
    \State \textbf{Case I:}
    \State Using Eqs.~\eqref{eq:case1} and ~\eqref{eq:log_dec}, we have
    \begin{equation}
      \sigma_{l}\rho = \frac{\sigma_{A}}{2} \otimes \sigma_{B}\rho',
    \end{equation}
    \State If $\sigma_{A}=I$, then $\sigma_{l} \mapsto \sigma_{B}.$ Otherwise, return zero.

    \Statex
    \State \textbf{Case II:}
    \State Using Eqs.~\eqref{eq:case2} and ~\eqref{eq:log_dec}, we have
    \begin{equation}
      \sigma_{l}\rho = \frac{1}{2}\bigl(\sigma_{A}\otimes\sigma_{B}\rho'
                       + \sigma_{A}\sigma_{1}\otimes\sigma_{B}g\rho'\bigr).
    \end{equation}
    \State If $\sigma_{A}=I$, then $\sigma_{l} \mapsto \sigma_{B}.$
    \State If $\sigma_{A}=\sigma_{1}$, then $\sigma_{l} \mapsto \sigma_{B}g.$ Otherwise, return zero.

    \Statex
    \State \textbf{Case III:}
    \State Using Eqs.~\eqref{eq:case3} and ~\eqref{eq:log_dec}, we have
    \begin{equation}
    \begin{aligned}
      \sigma_{l}\rho
        &= \tfrac12\bigl(
             \sigma_{A}\otimes\sigma_{B}\rho'
             + \sigma_{A}\sigma_{1}\otimes\sigma_{B}g_{1}\rho' \\
        &\qquad
             + \sigma_{A}\sigma_{2}\otimes\sigma_{B}g_{2}\rho'
             + \sigma_{A}\sigma_{1}\sigma_{2}\otimes\sigma_{B}g_{1}g_{2}\rho'
           \bigr).
    \end{aligned}
    \end{equation}
    \State If $\sigma_{A}=I$, then $\sigma_{l} \mapsto \sigma_{B}.$
    \State If $\sigma_{A}=\sigma_{1}$, then $\sigma_{l} \mapsto \sigma_{B}g_{1}.$
    \State If $\sigma_{A}=\sigma_{2}$, then $\sigma_{l} \mapsto \sigma_{B}g_{2}.$
    \State If $\sigma_{A}=\sigma_{1}\sigma_{2}$, then $\sigma_{l} \mapsto \sigma_{B}g_{1}g_{2}.$

    \Statex
    \State \textbf{Step 3:} Remove the first column in the tableau representation of $\rho$ and remove the first column in the vector representation of each $\sigma_{l}$.
\end{algorithmic}
\end{algorithm}
In Algorithm \ref{alg:update_algorithm}, we present an algorithm for updating the logical set $\mathcal{L}_{S}$ and the destabilizer set $\mathcal{D}$ when performing a measurement of the Pauli operator $P$ in Eq.~\eqref{eq:PauliOperator}. We now describe how to perform a measurement of the Pauli operator $P$, Eq.~\eqref{eq:PauliOperator}, with the projectors $(I\pm P)/2$, on the stabilizer state $\rho_{S}.$ After measurement, $\pm P$ becomes a stabilizer of $\rho_{S}$, and must therefore be to the stabilizer group. The list of stabilizer generators must then be updated so that they all commute with $P$. Furthermore, the list of destabilizers must be updated to preserve the required commutation relations.

\begin{figure}[htbp]
    \centering
    \includegraphics[width=3.4in]{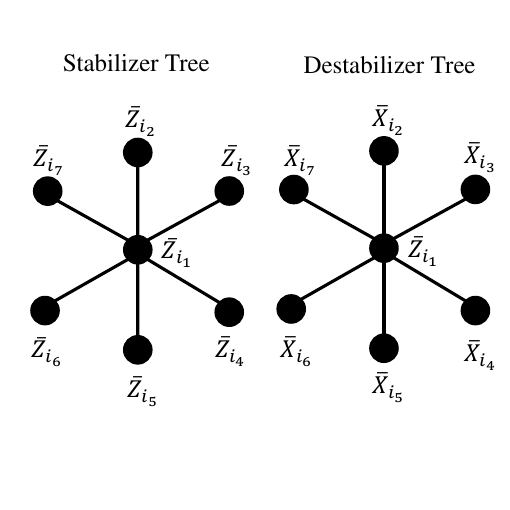}
    \caption{The stabilizer state $\rho_{S}$ has stabilizer generators $\{\overline{Z_{i_{1}}},\cdots,\overline{Z_{i_{r}}}\}$ and destabilizer generators $\{\overline{X_{i_{1}}},\cdots,\overline{X_{i_{r}}}\}$. We measure operator $g$ and have $\{g,\overline{Z_{i}}\}=0.$ Nodes represent anticommuting stabilizers, and vertices represent the choices of pairs to multiply. (Left) Example tree graph which depicts the update of the stabilizers after a measurement so that $[g,\bar{Z}_{i}] = 0$ $\forall i$. (Right) Example tree graph that depicts the update of the destabilizers after a measurement.}
    \label{fig:Tree}
\end{figure}

If a single stabilizer $\bar{g}_{i_{1}}$ anticommutes with $P$, we replace $\bar{g}_{i_{1}}$ with $\pm P$ and $\bar{d}_{i_{1}}$ with $\bar{g}_{i_{1}}$. When multiple stabilizers anticommute with $P$, we pair each anticommuting stabilizer and destabilizer into products and perform the updates $ \bar{g}_{i_{1}} \mapsto \bar{g}_{i_{1}} \cdot \bar{g}_{i_{k}}$, so the stabilizers now commute with $g_{i_{1}}$. The formation of a list of these products is equivalent to constructing a tree graph, shown in Fig.~\ref{fig:Tree}, where the nodes represent the anticommuting stabilizers and the edges represent the choices of pairs to multiply.
\begin{algorithm}[H]
\caption{Destabilizers and Logical Updates} \label{alg:update_algorithm}
\begin{algorithmic}
\State \textbf{Case I:} $[P,\mathcal{S}] \neq 0$
\State $P$ is not in the stabilizer group
\State \textbf{Step 1.} Identify the anticommuting stabilizers and their corresponding destabilizers: $\{g_{i_{1}}, \dots, g_{i_{k}}\}$ and $\{d_{i_{1}}, \dots, d_{i_{k}}\}.$
\State \textbf{Step 2.} Perform the update $g_{i_{j}} \mapsto g_{i_{1}} \cdot g_{i_{j}}$ and $d_{i_{j}} \mapsto g_{i_{1}} \cdot d_{i_{j}}$ for all $j \neq 1.$ This process is depicted graphically in Fig.~\ref{fig:Tree}.
\State \textbf{Step 3.} Form a list of anticommuting logicals $\{\ell_{i_{1}}^{x},\cdots,\ell_{i_{j}}^{x},\ell_{i_{1}}^{z},\cdots,\ell_{i_{k}}^{z}\} \subset \mathcal{L}_{S}.$ Here, either $i_{j}$ or $i_{k}$ could be $0$. Let $\ell_{i_{1}}^{x/z}$ denote the first anticommuting logical in the list. Update $\ell_{i} \rightarrow \ell_{i_{1}}^{x/z} \cdot \ell_{i}$ for all $\ell_{i} \in \mathcal{L}_{S}$ unless $\ell_{i}=\ell_{i_{1}}^{x/z}.$
\State \textbf{Step 4.} Add $\ell_{i_{1}}^{x/z}$ to $\mathcal{D}$ and $P$ to $\mathcal{S}$ so that $\ell_{i_{1}}^{x/z}$ and $P$ are anticommuting partners.
\State \textbf{Step 5.} Remove $\ell_{i_{1}}^{z/x}$, the anticommuting partner of $\ell_{i_{1}}^{x/z}.$

\State \textbf{Case II:} {$[g, s] = 0 \quad \forall s \in \mathcal{S}$ and $[g, \ell] = 0$ for all logicals $\ell \in \mathcal{L}$}
\State $g$ is in the stabilizer group
\State The outcome is determined by the sign of the product of the anticommuting destabilizers: $\pm g = \prod_{i=1}^{k} X_{i}$

\State \textbf{Case III:} {$[g, s] = 0 \quad \forall s \in \mathcal{S}$ and $\{g, \ell\} = 0$ for some logical $\ell \in \mathcal{L}$}
\State{Perform \textbf{Step 3} to \textbf{Step 5} of \textbf{Case I}.}
\end{algorithmic}
\end{algorithm}

We now present an algorithm for computing the R\'enyi entanglement entropy, which is valid only for the case $[\mathcal{L}_{\mathrm{LRSD}}, \mathcal{S}] = 0.$ In this case, we can write the density matrix $\rho$ as a sum over pairs of stabilizer states, where the states in each pair are orthogonal to each other. A partial trace over the environment is carried out for each stabilizer term. After tracing out the environment, we obtain a mixture of reduced density matrices. The purity, $\mathrm{tr}(\rho_{A}^{2})$, can be computed from overlaps of these density matrices using standard stabilizer techniques, from which $S_{n=2}$ can be easily computed. To compute the entanglement for arbitrary $n$, the reduced density matrix can be obtained from a partial trace over each term.

\begin{algorithm}[H]
\caption{R\'enyi Entanglement given $[\mathcal{L}_{\mathrm{LRSD}},\rho]=0$}
\begin{algorithmic}
    \State Let $A$ and $B$ be a bipartition of the system and
    \begin{equation}
    \rho = \sum_{\ell} \lambda_{\ell} \sigma_{\ell}\ketbra{\psi}.
    \end{equation}
    \State We compute $S_{n} (\rho)$ for some $n$ and rewrite $\rho$ as
\end{algorithmic}
\begin{equation}
\rho = \sum_{l} \lambda_{l} \sigma_{l} \ketbra{\psi}
\end{equation}
\begin{algorithmic}
    \Statex
    \State \textbf{Step 1:} Rewrite $\rho$ as:
\end{algorithmic}
\begin{equation}
\label{eq:rewrite_sum}
\rho = \sum_{l} \lambda_{l}\frac{1 + \sigma_{l}}{2} \ketbra{\psi} \frac{1 + \sigma_{l}}{2} - \lambda_{l}\frac{1 - \sigma_{l}}{2} \ketbra{\psi} \frac{1 - \sigma_{l}}{2}
\end{equation}
\begin{algorithmic}
    \State $\rho$ is now a sum of $2|\{\lambda_{\ell}\}|$ terms.
    \State \textbf{Step 2:} Perform a partial trace over region $B$ for each term in Eq.~\eqref{eq:rewrite_sum} using Algorithm \ref{alg:partial_trace_near_clifford}. Then, we rewrite
    \begin{equation}
    \label{eq:sum_add}
    \rho_{A} = \sum_{i=1}^{2|\{\lambda_{l}\}|} \lambda_{i} \rho_{A}^{(i)}.
    \end{equation}
\State \textbf{Case I:} For $n=2$, we compute the purity
\begin{equation}
\mathrm{tr}(\rho_{A}^{2}) = \sum_{i,j} \lambda_{i} \lambda_{j} \mathrm{tr}(\rho_{A}^{(i)} \rho_{A}^{(j)})
\end{equation}
We can rewrite each trace term in the sum as:
\begin{equation}
\label{eq:trace_successive}
\mathrm{tr}[\prod_{k=1}^{r} \frac{I+(-1)^{r_{k}} g_{k}}{2} \rho_{A}^{(j)}],
\end{equation}
where each $g_{k}$ is a stabilizer of $\rho_{A}^{(i)}$ and $r_{i} \in \{0,1\}$ defines the phase. Eq.~\eqref{eq:trace_successive} is the probability of successively measuring the $r_{i}$ stabilizers of $\rho_{A}^{(i)}$ with the outcomes $\pm 1$ encoded via the $\{r_{k}\}.$ If $\rho_{A}^{(i)}$ and $\rho_{A}^{(j)}$ are pure states, then $\rho_{A}^{(i)}=\ketbra{\psi_{i}}$ and $\rho_{A}^{(j)}=\ketbra{\psi_{j}}.$ Then, $\mathrm{tr}[\rho_{A}^{(i)}\rho_{A}^{(j)}]=\abs{\braket{\psi_{i}}{\psi_{j}}}^{2}$ can be computed using the inner product formula in Ref.\cite{aaronson2004improved}. Namely, enumerate the stabilizer groups $\mathcal{S}_{\psi_{i}}$ and $\mathcal{S}_{\psi_{j}}$ for $\ket{\psi_{i}}$ and $\ket{\psi_{j}}.$ Let
\begin{equation}
s = |\mathcal{S}_{\psi_{i}} \cap \mathcal{S}_{\psi_{j}}|
\end{equation}
be the number of stabilizer generators the two states have in common. Then,
\begin{equation}
  \innerproduct{\psi_{i}}{\psi_{j}} =
    \begin{cases}
      1 & \text{if $s \geq L$}\\
      0 & \text{if $\mathcal{S}_{\psi_{i}}$ and $\mathcal{S}_{\psi_{j}}$ contain the same stabilizer}\\
        & \text{generator with opposite signs}\\
      2^{-\frac{s}{2}} & \text{otherwise}
    \end{cases}
\end{equation}

\State \textbf{Case II:} For arbitrary $n$, each term in Eq.~\eqref{eq:sum_add} can be represented as a matrix on the full Hilbert space, and thus $\rho_{A}$ can be constructed explicitly. The entanglement can then be computed directly from $\rho_{A}$.

\end{algorithmic}

\end{algorithm}

\subsection{Bell Sampling and Magic Computation Algorithms}
\label{sec:BellSample}

\subsubsection{Bell sampling}
\label{sec:BellSampling}
\begin{figure}[htbp]
    \includegraphics[width=3.4in]{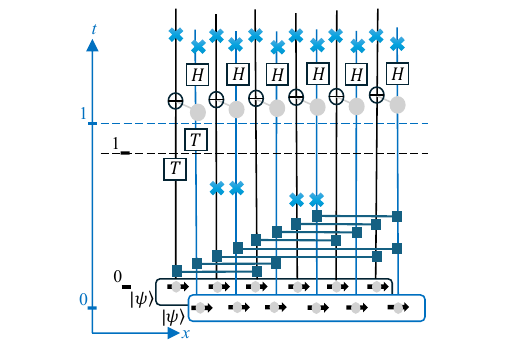}
    \caption{
    Bell sampling setup for the single-pair all-to-all model. The quantum circuit to measure the magic $\mathcal{M}$ of an $L$-qubit state $\ket{\psi} = K\ket{+x}^{\otimes N}$ where $K$ is the Kraus operator of the circuit gates and measurements. We prepare two copies $\ket{\psi} \otimes \ket{\psi}$, and perform a measurement in the Bell basis after the final time step.
    }
    \label{fig:bellsamplev2}
\end{figure}
To compute Eq.~\eqref{eq:magic_equation_main},
we need to Bell sample the Pauli strings from the state $\ket{\psi}$, as illustrated in Fig.~\ref{fig:bellsamplev2}. To define the Bell samples, we first define the Pauli matrices $\sigma_{00} = I$, $\sigma_{01} = X$, $\sigma_{10} = Z$, and $\sigma_{11} = Y$. The $4^{L}$ Pauli strings are $L$-qubit tensor products of Pauli matrices, which can be written as $\sigma_{\bm{r}} = \bigotimes_{j=1}^{L} \sigma_{r_{j}, r_{j+L}}$, where $\bm{r} \in \{0,1\}^{2L}$.
We first prepare two copies of the state, $\ket{\Psi}=\ket{\psi}\otimes\ket{\psi}$, and then measure pairs of qubits $(j,j+L)$ in the Bell basis, which is defined as,
\begin{equation}
\begin{array}{cc}
\ket{\Phi^+} = \dfrac{1}{\sqrt{2}} (\ket{00} + \ket{11}), &
\ket{\Psi^+} = \dfrac{1}{\sqrt{2}} (\ket{01} + \ket{10}) \\
\ket{\Phi^-} = \dfrac{1}{\sqrt{2}} (\ket{00} - \ket{11}), &
\ket{\Psi^-} = \dfrac{1}{\sqrt{2}} (\ket{01} - \ket{10})
\end{array}
\end{equation}

Following Ref. \cite{lai2022learning},  the probability of obtaining the sequence of outcomes $\vec{r}$ is:
\begin{equation}
\label{eq:probability}
p({\vec{r}}) = 2^{-L}\abs{\mel{\psi}{\sigma_{\vec{r}}}{\psi^{*}}}^{2},
\end{equation}
where $\ket{\psi^{*}}$ is the complex conjugate of $\ket{\psi}$ ~\cite{montanaro2017learning}.
For a stabilizer state $\ket{\psi}$ with stabilizer group $\mathcal{S} \subset \mathcal{P}_{L}$, there exists a unique Pauli operator $g$ such that $g\ket{\psi}=\ket{\psi^{*}}.$ $p(\vec{r})$ is non-zero if and only if $\sigma_{\vec{r}}$ belongs to a single coset $g\mathcal{S}$ of the stabilizer group $\mathcal{S}$ of $\ket{\psi}.$ Bell sampling produces a uniform distribution over all $2^{L}$ elements of the coset. All operators inside that coset are sampled with equal probability $1/2^{L}$, while those outside have zero probability. The Bell samples lie in a vector space of dimension $G'=L$, so $\mathcal{M}=0$ for stabilizer states. For a non-stabilizer pure state, the Bell samples are not in a single coset. The Bell samples form an unstructured subset of $\mathcal{P}_{L}$, which has dimension $G'>L$, so $\mathcal{M'} >0$ for non-stabilizer states.

We develop a Monte Carlo method for Bell sampling from the distribution $p(\vec{r}),$ in Appendix~\ref{sec:BellSample}, building on methods introduced in Ref.~\cite{lai2022learning}. Bell sampling can be intractable for large $L$ as the number of $\sigma_{\vec{r}}$ for which $p(\vec{r})=0$ is exponentially growing in $L$. Our method always generates samples $\sigma_{\vec{r}}$ with $p({\vec{r}}) \neq 0$ at each sampling step. A sampling step means drawing one independent sample from $p(\vec{r})$. We compute $\mathcal{M}$ after each sampling step, and find that $\mathcal{M}$ converges in fewer than $\tau < 2L$ sampling steps for all measurement rates, provided  $L \leq 256$ and $0 < \beta < 1$, at $T$-gate rates given by $p_T = 12 / L^{\beta}$, as shown in Appendix \ref{sec:BellSample}. We also find that our Monte Carlo sampling method converges to the exact value of the magic obtained by exhausting all possible $4^{L}$ samples in polynomial time. A reason we use the stabilizer nullity to quantify magic is that it admits an exactly correct result in a very low number of samples, shown in \ref{sec:BellSample}.

\subsubsection{Bell Sampling Algorithm}
We now introduce a Monte Carlo sampling algorithm that efficiently produces Bell samples for a specific subclass of states that include, but are not limited to, stabilizer states. The sampler only visits Bell samples, $\sigma_{\vec{r}}$, which have nonzero probabilities $p(\vec{r}) > 0$, where $p(\vec{r})$ is defined in Eq. ~\eqref{eq:probability}. This algorithm is adapted from Ref. \cite{lai2022learning}. The non-stabilizer state must be able to be written as $\ket{\psi}=\mathcal{T} \mathcal{C}\ket{\psi_{o}}$, where $\ket{\psi_{o}}$ is a stabilizer state, $\mathcal{C}$ is a Clifford circuit, and $\mathcal{T}$ is a layer of $T$-gates. The non-Clifford layer factorizes as
\begin{equation}
\label{eq:single_t_layer}
\mathcal{T} = \bigotimes_{i=1}^{L} T^{N_{i}},
\end{equation}
where $N_{i}$ is the number of $T$-gates applied to qubit $i.$

Using $T^{2}=S$, $T^{4}=Z$, and $T^{8}=I$, each power $T^{N_{i}}$ can be decomposed into a product of Clifford and non-Clifford gates:
\begin{equation}
\label{eq:T_layer}
\begin{aligned}
T^{N_i} &= T^{a_i} S^{b_i} Z^{c_i}, \quad
a_i = N_i \bmod 2, \quad
b_i = \left\lfloor \frac{N_i \bmod 4}{2} \right\rfloor, \\
c_i &= \left\lfloor \frac{N_i}{4} \right\rfloor.
\end{aligned}
\end{equation}
Here, $\left\lfloor \cdots \right\rfloor$ is the floor function and $S=\mathrm{diag}(1,i).$ The explicit Clifford expansion allows at most one residual non-Clifford gate to act on each qubit. The rest of the $T$-gates are absorbed into the Clifford part of the circuit, so the number of $T$-gates remaining is $k = \sum_{i=1}^{n} a_{i}.$

We now consider how the density matrix $\rho_{S}$, for the initial state $\mathcal{C}\ket{\psi_{o}}$, evolves under the $T$-gate layer in Eq.~\eqref{eq:single_t_layer}. The final state $\rho$ can be represented as a product of a mixed stabilizer state and a product over $k$ terms, each of which is written as a sum over three projectors:
\begin{equation}
\label{eq:DensityMatrixMagic}
\begin{aligned}
\rho
&= \prod_{i=1}^{n-k} \left( \frac{I + g_{i}}{2} \right)
\left\{ \prod_{i=n-k+1}^{n} \left(
\alpha_{1} \left[ \frac{I + g_{i}}{2} \right]
+ \alpha_{2} \left[ \frac{I - g_{i}}{2} \right] \right. \right. \\
&\qquad\left. \left.
+ \alpha_{3} \left[ \frac{I + h_{i}}{2} \right]
\right) \right\}
\end{aligned}
\end{equation}
where $\alpha_{1} = \frac{1}{2}$, $\alpha_{2} = \frac{1-\sqrt{2}}{2}$, $\alpha_{3} = \frac{\sqrt{2}}{2}$. Expanding the product over the $k$ stabilizer triplets yields a weighted sum of $3^{k}$ stabilizer states. The $L-k$ operators $g_{1},\cdots,g_{L-k}$ stabilize all of the stabilizer states. We call these isotropic stabilizers using the terminology in Ref.~\cite{lai2022learning}. Along with a set of isotropic destabilizers $d_{1},\cdots,d_{k}$, they define a mixed stabilizer state. There are additional $k$ triplets of stabilizer generators $(g_{L-k+1},-g_{L-k+1},h_{L-k+1}), \cdots, (g_{L},-g_{L},h_{L}).$ Each of these generators is called a \textit{primary symplectic stabilizer} (PSS) following Ref.~\cite{lai2022learning}. Each PSS stabilizes $1/3$ of the stabilizer states. We require the commutation relations $[g_{i},g_{j}]=0, [h_{i},h_{j}]=0, \{g_{i},h_{j}\}=0$ for $i=j$ and $[g_{i},h_{j}]=0$ for $i \neq j$. The state is written in this form, as Bell sampling can be performed efficiently. The Bell samples can be quickly produced from products of the isotropic stabilizers and PSS following Algorithm \ref{alg:BellSamplingZbasis}.

\subsubsection{Magic Computation}
In this section, we explain how to compute the magic of a circuit using the representation in Eq.~\eqref{eq:DensityMatrixMagic}. The key idea is to use the stabilizer tableau, a compact binary encoding of the stabilizers, to bring the stabilizer generators into a compact form as in Eq.~\eqref{eq:CanonicalForm}. The procedure to compute the magic is summarized in Algorithm~\ref{alg:magic_compute}.
We recall the explicit form of the stabilizer tableau for $\rho_{S}.$ We first recall again that a Pauli operator $P$ can be expressed as a $2L$ binary vector $(\mathbf{x}|\mathbf{z})$ as discussed at the beginning of Section \ref{sec:magic}. The state $\rho_{S}$ has generators $\{g_{1},\cdots,g_{L}\},$ where $g_{i} = c_{i}P_{i}$ and $c_{i} = \pm 1.$ Then, the tableau is
\[
\left(
\begin{array}{cccc|cccc|c}
x_{11} & x_{12} & \cdots & x_{1L}  & z_{11} & z_{12} & \cdots & z_{1L}  & \operatorname{sgn}(c_{1}) \\[2pt]
x_{21} & x_{22} & \cdots & x_{2L}  & z_{21} & z_{22} & \cdots & z_{2L}  & \operatorname{sgn}(c_{2}) \\[2pt]
\vdots & \vdots & \ddots & \vdots  & \vdots & \vdots & \ddots & \vdots  & \vdots                   \\[2pt]
x_{L1} & x_{L2} & \cdots & x_{LL}  & z_{L1} & z_{L2} & \cdots & z_{LL}  & \operatorname{sgn}(c_{L})
\end{array}
\right)
\]
where $\mathrm{sgn}(c_{i}) = \frac{1-c_{i}}{2}.$ This can be more compactly represented as: $[X|Z|\vec{c}]$, where $X,Z \in \{0,1\}^{L \times L}$ and $\vec{c} \in \{0,1\}^{L}$.

\begin{algorithm}[H]
\caption{Compute Magic}
\label{alg:magic_compute}
\begin{algorithmic}
    \State \textbf{Step 1:} Run the Clifford circuit $U_{c}$ to obtain $[X|Z|\vec{c}]$
\end{algorithmic}
\begin{algorithmic}
    \State \textbf{Step 2:} Perform row operations $g_{i} \rightarrow g_{i} \cdot g_{j}$ and row swaps $g_{i} \leftrightarrow g_{j}$ such that $X=I_{L \times L}$. Each stabilizer is then:
\end{algorithmic}
\begin{equation}
\label{eq:CanonicalForm}
g_{j} = \bigotimes_{i=1}^{L} X_{i}^{\delta_{ij}} Z^{f(i,j)},
\end{equation}
where $f(i,j) \in \{0,1\}.$ Here, $\delta_{ij}$ is the Kronecker delta.
\begin{algorithmic}
    \State \textbf{Step 3:} Apply Algorithm \ref{alg:bell_sample_complex} on $\ket{\psi} \otimes \ket{\psi^{*}}$ to obtain samples $\sigma_{1},\cdots,\sigma_{n}$
\end{algorithmic}
\begin{algorithmic}
    \State \textbf{Step 4:} Apply Algorithm \ref {alg:magic_estimation} to compute $\mathcal{M}$.
\end{algorithmic}
\end{algorithm}
After generating samples $\sigma_{1},\cdots,\sigma_{\mathcal{N}_{s}}$ from the Bell measurement procedure in Step 3 of Algorithm \ref{alg:magic_compute}, we can proceed to estimate the stabilizer nullity $\mathcal{M}$ of the state. This estimation is performed by computing the rank of pairwise differences between the strings.
\begin{algorithm}[H]
\caption{Magic Estimation}
\label{alg:magic_estimation}
\begin{algorithmic}
    \State \textbf{Input:} Number of sampling steps $\tau$ and input state $\ket{\psi}$
    \State \textbf{Step 1:} Collect a Bell sample from $\ket{\psi}$ each of the $\tau$ times, obtaining unique Pauli strings $\sigma_{1},\sigma_{2},\cdots,\sigma_{\mathcal{N}_{s}}.$ Here, $\tau \geq \mathcal{N}_{s}$ since two sampling steps may yield the same sample.
    \State \textbf{Step 2:} Compute all Bell differences $\sigma_{(i,j)}=\sigma_{i}\oplus\sigma_{j}$, where $\oplus$ is the XOR operation on the binary representations of the Pauli strings.
    \State \textbf{Step 3:} Define
    \begin{equation}
    G' = \mathrm{rank}[\mathrm{span}(\{\sigma_{(i,j)}\})]
    \end{equation}
    and the stabilizer nullity
    \begin{equation}
    \mathcal{M} = G' - L
    \label{eq:magic_definition}
    \end{equation}
\end{algorithmic}
\end{algorithm}

We now present an algorithm for Bell sampling from the state in Eq.~\eqref{eq:DensityMatrixMagic}. Recall from Section \ref{sec:BellSampling} that a Bell sample $\sigma_{\vec{r}}$ is obtained from $\ket{\psi} \otimes \ket{\psi}$ with probability $p(\vec{r})$. This is equivalent to measuring each qubit pair in the Bell basis and obtaining the outcome sequence $\vec{r}$ with Born probability $p(\vec{r})$. We now define $\sigma_{\vec{r}_{o}}$ such that
\begin{equation}
\label{eq:sigma_ro}
\sigma_{\vec{r}_{o}}\ket{\psi} = \ket{\psi^{*}},
\end{equation}
and
\begin{equation}
\label{eq:psiprob}
p(\vec{r}) = \frac{\mel{\psi}{\sigma_{\vec{r}}}{\psi^{*}}}{2^{L}} = \frac{\mel{\psi}{\sigma_{\vec{r}\oplus \vec{r}_{o}}}{\psi}}{2^{L}}.
\end{equation}
We recall that a Bell sample $\sigma_{\vec{r}}$ from $\ket{\psi} \otimes \ket{\psi^{*}}$ occurs with probability~\cite{montanaro2017learning}
\begin{equation}
\label{eq:psiconjprob}
p(\vec{r}) = \frac{\abs{\mel{\psi}{\sigma_{\vec{r}}}{\psi}}^{2}}{2^{L}}.
\end{equation}
From Equations~\eqref{eq:psiprob} and ~\eqref{eq:psiconjprob} we see that obtaining a Bell sample $\sigma_{\vec{r}}$ from $\ket{\psi} \otimes \ket{\psi}$ is equivalent to obtaining a Bell sample $\sigma_{\vec{r'}}$ from $\ket{\psi} \otimes \ket{\psi^{*}}$, and then shifting it by $\vec{r}_{o}$: $\sigma_{\vec{r}}=\sigma_{\vec{r'}\oplus\vec{r_{o}}}.$ We then present in Algorithm \ref{alg:bell_sample_complex} how to compute $\sigma_{\vec{r}_{o}}$ and then in Algorithm \ref{alg:bell_sample_real} how to obtain Bell samples from $\ket{\psi} \otimes \ket{\psi^{*}}.$

\begin{algorithm}[H]
\caption{Bell Sampling from $\ket{\psi} \otimes \ket{\psi}$}
\label{alg:bell_sample_complex}
\begin{algorithmic}
    \State \textbf{Input:} Isotropic stabilizers $\{g_{1},\dots,g_{L-k}\}$, isotropic destabilizers $\{h_{1},\dots,h_{L-k}\}$, PSS stabilizers
    $\{g_{L-k+1},\dots,g_{L}\}$ and
    $\{h_{L-k+1},\dots,h_{L}\}$; desired number of samples $n$.
    \Statex

    \State \textbf{Step 1:} Obtain samples $\sigma_{1},\dots,\sigma_{n}$
           using Algorithm \ref{alg:bell_sample_real} on  $\ket{\psi} \otimes \ket{\psi}$.
    \State \textbf{Step 2:} Compute $\sigma_{\vec{r}_{o}}$, as defined in Eq. ~\eqref{eq:sigma_ro},

    \State \textbf{Case I:} $g_{i}^{*}=-g_{i}$ for some $i$, where $g_{i}^{*}$ is the complex conjugate of $g_{i}$.
\State \textbf{Step 1:} Ensure $[h_{i},h_{j}]=0$. If $\{h_{i},h_{j}\}=0$, set $h_{i}\mapsto h_{i}g_{j}$ for $i\in\{1,\dots,k\}$ and $j\in\{L-k+1,\dots,L\}$.
\State \textbf{Step 2:} $\sigma_{\vec{r_{o}}} = h_{1}$.
\State \textbf{Step 3:} Ensure $g_{i}^{*}=g_{i}$ for $i=2,\dots,L-k$ and $\{g_{i},h_{i}\}=0$. For all $g_{i}^{*}=-g_{i}$, set $\sigma_{\vec{r_{o}}}\mapsto\sigma_{\vec{r_{o}}}h_{i}$ and $g_{i}\mapsto g_{i}g_{1}$.
\State \textbf{Step 4:} Ensure $g_{i}^{*}=g_{i}$ and $h_{i}^{*}=h_{i}$ for $i=L-k+1,\dots,L$. For all $h_{i}^{*}=-h_{i}$, set $h_{i} \mapsto g_{1}h_{i}$ and $\sigma_{\vec{r_{o}}}\mapsto\sigma_{\vec{r_{o}}}g_{i}$; for all $g_{i}^{*}=-g_{i}$, set $g_{i}\mapsto g_{1}g_{i}$ and $\sigma_{\vec{r_{o}}}\mapsto\sigma_{\vec{r_{o}}}h_{i}$.
\State \textbf{Case II:} $g_{i}^{*}=g_{i}$ for all $1\leq i \leq L-k$
\State Define $\mathcal{G}=\{i\mid L-k+1\le i\le L, g_{i}^{*}=-g_{i}\}$
\State Define $\mathcal{H}=\{i\mid L-k+1\le i\le L, h_{i}^{*}=-h_{i}\}$
\State Let
$\displaystyle \sigma_{\vec{r}_{o}}=
\Bigl(\prod_{i\in\mathcal{G}}h_{i}\Bigr)
\Bigl(\prod_{l\in\mathcal{H}}g_{l}\Bigr)$
\State \textbf{Step 3:} Return $\sigma_{1 \oplus \vec{r}_{o}}, \cdots, \sigma_{N_{s} \oplus \vec{r}_{o}}$
\end{algorithmic}
\end{algorithm}

We recall from Ref.~\cite{lai2022learning} that, if $\pm\sigma_{r}$ is an
\emph{isotropic stabilizer}, then
\begin{equation}\label{eq:single-stab-prob}
  P_{\mathrm{sample}}\bigl(\pm\sigma_{r}\bigr)=\frac{1}{2^{L}},
\end{equation}
where $L$ is the number of qubits.

If $\pm\sigma_{r}$ is a product of $m$ Pauli-string stabilizers (an
\emph{$m$-PSS}), the probability of sampling \emph{any} $m$-PSS is
\begin{equation}\label{eq:mpss-prob-def}
  P_{\mathrm{sample}}\bigl(m\text{-PSS}\bigr)
  =
  \frac{n_{m\text{-PSS}}}{2^{L}},
\end{equation}
where $n_{m\text{-PSS}}$ denotes the total number of such objects.

There are $k$ disjoint PSS \emph{pairs}.
To form an $m$-PSS, we
(i) choose the $m$ pairs that actually appear (giving the factor
$\binom{k}{m}$),
(ii) choose \emph{one} stabilizer from each selected pair
($g_{n-k+i}$ or $h_{n-k+i}$, giving $2^{m}$ choices), and
(iii) multiply by the $2^{n-k}$ isotropic stabilizers that commute with
all pairs.  Thus
\begin{equation}\label{eq:mpss-count}
  n_{m\text{-PSS}} =
  2^{m}\binom{k}{m}2^{n-k}.
\end{equation}

Substituting~\eqref{eq:mpss-count} into~\eqref{eq:mpss-prob-def} and
using~\eqref{eq:single-stab-prob} gives
\begin{equation}\label{eq:mpss-prob-final}
  P_{\mathrm{sample}}\bigl(m\text{-PSS}\bigr)
  =
  2^{-k}\binom{k}{m}.
\end{equation}

\begin{algorithm}[H]
\caption{Bell Sampling from $\ket{\psi} \otimes \ket{\psi^{*}}$}
\label{alg:bell_sample_real}
\begin{algorithmic}
    \State \textbf{Input:} Isotropic stabilizers
           $\{g_{1},\dots,g_{L-k}\}$, PSS stabilizers
           $\{g_{L-k+1},\dots,g_{L}\}$ and
           $\{h_{L-k+1},\dots,h_{L}\}$; desired number of samples $n$.
    \Statex

    \State \textbf{Step 1:} Generate a random isotropic stabilizer
          \begin{equation}
             \sigma_{\mathrm{iso}} = \prod_{i=1}^{L-k} g_{i}^{f(i)},
          \end{equation}
          where each $f(i)\in\{0,1\}$ with equal probability.
    \Statex

    \State \textbf{Step 2:} Generate a random product of PSS
          \begin{equation}
             \sigma_{\mathrm{PSS}} = \prod_{j=1}^{m}
             g_{L-k+i_{j}}^{g(i_{j})}
             h_{L-k+i_{j}}^{1-g(i_{j})},
          \end{equation}
          where $m\in[0,k]$ is chosen according to the $P(m) = 2^{-k}\binom{k}{m}$ (Eq.~\eqref{eq:mpss-prob-final}).
    \Statex

    \State \textbf{Step 3:} Collect the Bell sample $\sigma_{\mathrm{iso}}\sigma_{\mathrm{PSS}}$
    \Statex

    \State \textbf{Step 4:} Repeat Steps 1–3 until $n$ distinct samples $\sigma_{\vec{r}_{1}},\cdots,\sigma_{\vec{r}_{n}}$are obtained.
    \Statex

\end{algorithmic}
\label{alg:BellSamplingZbasis}
\end{algorithm}

\begin{figure}[htbp]
    \includegraphics[width=3.4in]{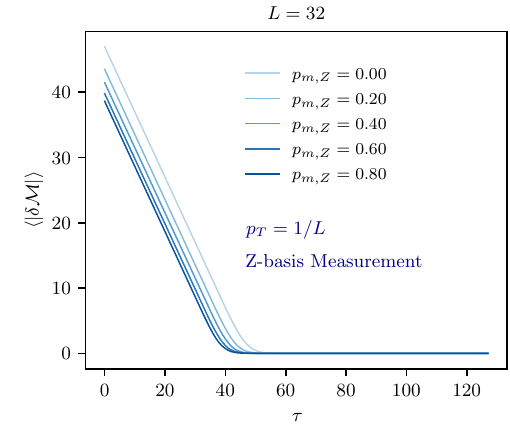}
    \caption{Absolute value of the difference in the exact stabilizer nullity, defined in Eq.~\eqref{eq:magic_definition}, and the numerical estimate, $\abs{\delta \overline{\mathcal{M}}}$ as a function of the sampling time $\tau$. $\tau$ is defined in Algorithm \ref{alg:bell_sample_real}. $\overline{\cdots}$ denotes averaging over circuit trajectories. The exact $\mathcal{M}$ for each circuit is obtained by sampling all possible Pauli operators to compute Eq.~\eqref{eq:magic_definition}, and the numerical estimate is obtained using $\tau$ sampling steps of Algorithm \ref{alg:bell_sample_real}. Data is obtained in the single-pair all-to-all model with $Z$-basis measurements, system size $L=48$, and measurement probabilities $p_{m} \in \{0.0,\ 0.2,\ 0.4,\ 0.6,\ 0.8\}$. We average over $\sim10^{3}$ iterations to obtain each curve. For each fixed $p_{m}$ and circuit, the locations of unitary gates and measurements are the same for the numerical and exact estimates of $\mathcal{M}$. }
    \label{fig:mag_sat_time_check}
\end{figure}

To test the accuracy of Algorithm \ref{alg:bell_sample_real}, we fix the location of the unitary gates and measurement locations for many circuits. We sample all $4^{L}$ possible Pauli operators on $L$ qubits. We then compute $\mathcal{M}_{\mathrm{exact}}$ using Eq.~\eqref{eq:magic_definition} with these samples. We also obtain a numerical estimate $\mathcal{M}_{\mathrm{numerical}}$ for the same set of circuits, using Algorithm \ref{alg:bell_sample_real} with $\tau$ samples. Then, define $|\delta \mathcal{M}|= \mathcal{M}_{\mathrm{numerical}}-\mathcal{M}_{\mathrm{exact}}.$ The absolute value $|\delta \mathcal{M}|$ will then be positive and decrease to 0 with increasing $\tau$, as shown in Fig.~\ref{fig:mag_sat_time_check}.

We now demonstrate our algorithm on the single-pair all-to-all model with $Z$-basis measurements. From the discussion at the beginning of Section \ref{sec:magic_complexity}, we observe that the state can be factorized with the $T$-gate layer in Eq. ~\eqref{eq:single_t_layer} acting at the final timestep.

\begin{figure*}[tbp]
    \centering
    \includegraphics[width=7in]{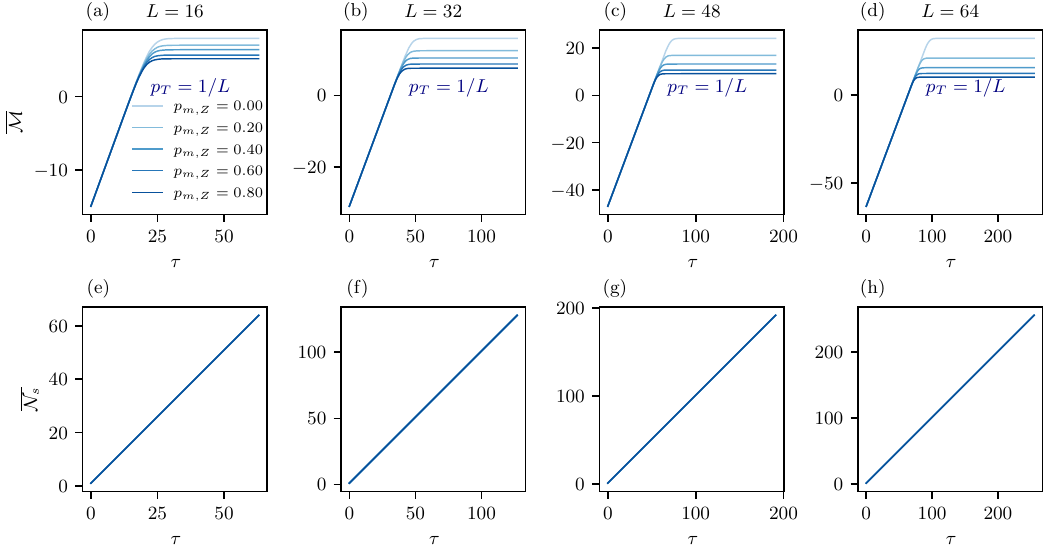}
    \caption{
        The top row shows the stabilizer nullity $\mathcal{M}$, defined in Eq.~\eqref{eq:magic_definition}, as a function of the sampling time $\tau$ for fixed measurement probabilities $p_{m} \in \{0.0, 0.2, 0.4, 0.6, 0.8\}$. $\overline{\cdots}$ denotes averaging over circuit trajectories. We use Algorithm \ref{alg:bell_sample_real} to compute $\mathcal{M}$, where $\tau$ is defined in \ref{alg:bell_sample_real}. Data is obtained from the single-pair all-to-all model with $X$-basis measurements. Each curve represents an average over $\sim 10^{3}$ circuit realizations. Results are shown for system sizes: (a) $L = 16$, (b) $L = 32$, (c) $L = 48$, and (d) $L = 64$.
        The bottom row, panels (e-h), shows the number of distinct Bell samples $\mathcal{N}_{s}$, defined in Step 1 of Algorithm \ref{alg:bell_sample_real}, as a function of the sampling time $\tau$. Data was obtained using the same set of circuits as in the top row, and averaged over $\sim 10^{3}$ trajectories.
    }
    \label{fig:mag_sat_time}
\end{figure*}

Fig.~\ref{fig:mag_sat_time} illustrates the convergence behavior of the stabilizer nullity $\mathcal{M}$ and the number of distinct Bell samples $\mathcal{N}_{s}$ as a function of the number of samples $\tau$, for different measurement rates $p_{m}.$ In the top row, we observe that $\mathcal{M}$ exhibits rapid growth followed by saturation for all values of $p_{m}.$ The saturation value decreases as $p_{m}$ increases, reflecting expected suppression of non-stabilizerness by projective measurements. Additionally, saturation occurs earlier for larger $p_{m}$, which correlates with the reduced complexity of the state. The bottom row, $\mathcal{N}_{s}$, the number of distinct Bell samples obtained during sampling, is independent of the measurement rate $p_{m}.$ It grows at a rate of approximately one per sampling step, indicating that it is rare for the same Pauli to be sampled twice in the same Monte Carlo run.

\clearpage
\section{Purification, cluster, and all-to-all entanglement transitions} \label{sec:EntanglementTransitions}

\subsection{Cluster}
\label{sec:ClusterTransition}
We begin with the connectivity transition, which is a percolation transition from a connected state (when all qubits are entangled) to a disconnected state (when the state effectively breaks into multiple disentangled clusters). The quantum state factorizes as
\begin{equation}\label{eq:cluster_factorization}
    \ket{\psi}=\bigotimes_{i}\ket{\psi_{C_{i}} } ,
\end{equation}
where each $C_{i}$ denotes a disjoint subset of qubits, or cluster, and $\ket{\psi_{C_{i}}}$ is an entangled pure state supported on that cluster. The full state $\ket{\psi}$ is a product across these clusters with no entanglement between disjoint $C_{i}$.

Therefore, we can probe the critical connectivity (cc) measurement rate $p_{m}^{cc}$ by tracking the trajectory-averaged number of qubits in the largest cluster $n_{\text{max}}\equiv\abs{C_{\text{max}}}$.
Inside the connected phase, the size should be $\mathcal{O}(L)$, as all qubits are entangled, whereas in the disconnected phase, the number should be $\mathcal{O}(1)$.
Numerically, we convert each stabilizer code into an equivalent graph state to identify entangled clusters efficiently \cite{nest2004efficient,anders2006fast}. This graph-state conversion is only valid for stabilizer states; when non-Clifford gates such as $T$-gates are present, it is not yet known how an analogous mapping can be performed.
Because every stabilizer state is local‑Clifford equivalent to a graph state, we first apply single‑qubit Clifford rotations that bring the stabilizer generators $g_{i}$ into the canonical graph form:
\begin{equation}
g_i = X_i \bigotimes_{j \in \mathcal N(i)} Z_j,
\end{equation}
where $\mathcal N(i)$ is the neighbor set of qubit $i$.
We then define a graph $G=(V,E)$ with one vertex per qubit, $V=\{1,\dots,L\}$, and an edge $(i,j)\in E$ whenever $j\in\mathcal N(i)$; equivalently, each edge corresponds to a $\mathrm{CZ}$ gate in the standard preparation circuit:

\begin{equation}
\ket{G} = \prod_{(i,j)\in E} \mathrm{CZ}_{ij} \ket{+}^{\otimes L}.
\end{equation}

The connected components of $G$ give precisely the entangled clusters $C_i$.

We study the connectivity transition in the single-pair all-to-all model with both $X$-basis and $Z$-basis measurements. The probability of performing a measurement in the $X$-basis given a measurement is performed is denoted by $p_{x \vee z}$. We also turn off the $T$-gates in the circuit, $p_T=0$.
The numerical phase diagram is shown in Fig.~\ref{fig:cluster_pd}, where we find that the estimated phase boundary between disconnected phase and connected phase is roughly always around $p_{m}^{cc} \approx 0.66(1)$, independent of $x$-basis measurement $p_{m,x}$.

\begin{figure*}[htbp]
    \centering
    \includegraphics[width=7in]{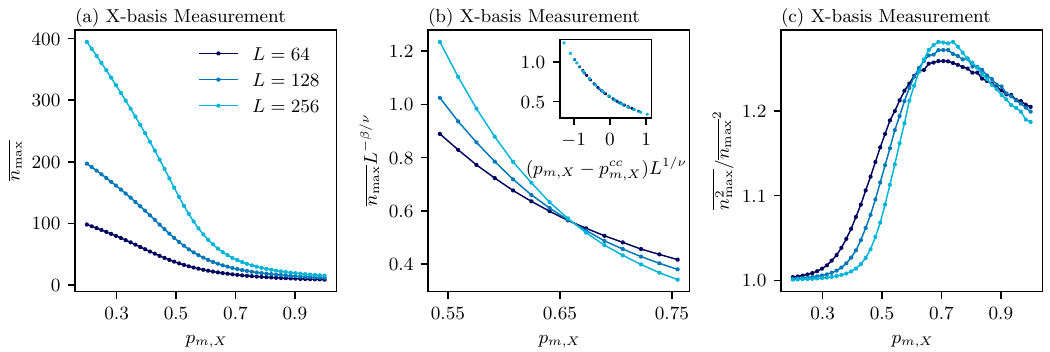}
    \caption{
(a) The first moment of the maximal cluster size $\overline{n_{\mathrm{max}}}$ as a function of the measurement probability $p_{m}$. $\overline{\cdots}$ denotes averaging over circuit trajectories. Data is shown for the single-pair all-to-all model with $X$-basis measurements.
(b) Rescaled first moment of the maximal cluster size $\overline{n_{\mathrm{max}}} L^{-\beta/\nu}$ as a function of the measurement probability $p_{m}$, which exhibits a crossing at the critical connectivity (cc) rate $p_{m}=p_{m}^{cc}$ (Inset) Scaling collapse for maximal cluster size used to obtain the critical exponents in Table \ref{tab:single_layer_cluster}. (c) Normalized variance of the peak cluster size, $\overline{n_{\mathrm{max}}^{2}}/\overline{n_{\mathrm{max}}}^{2}$ as a function of measurement probability $p_{m}$. The location of the peak is a proxy of $p_{m}^{cc}$. }
    \label{fig:three_panel_cluster}
\end{figure*}

\begin{figure} \label{fig:cluster_pd}
\includegraphics[width=3.4in]{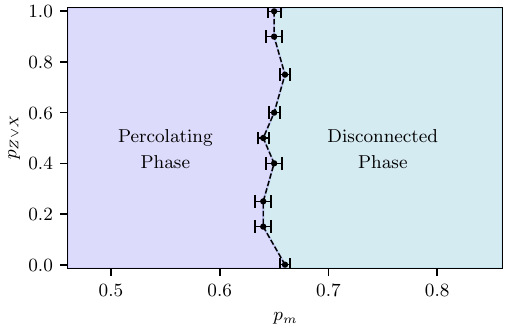}
\caption{Phase diagram for the connectivity transition in the single-pair all-to-all model (see Fig.~\ref{fig:circuit-schematic}(a)) as a function of measurement probability $p_{m}$ and probability of the measurements in $X$-basis, given a measurement, $p_{X \vee Z}$. The system undergoes a transition from a connected phase to a disconnected phase at the critical connectivity (cc) rate $p_{m}^{cc} = 0.66(1)$, beyond which the wave function becomes a product state over multiple clusters.}
\end{figure}

This phase boundary of the connectivity transition is detected by directly estimating the number of qubits in the largest cluster defined as \cite{gullans_huse_pc_2023}
\begin{equation}\label{eq:clustersize}
n_{\text{max}} = \max_i |C_i|,
\end{equation}
where $C_i$ represents the $i$-th cluster and $|C_i|$ is the number of qubits in that cluster. As shown in Fig.~\ref{fig:three_panel_cluster}(b), we find that the trajectory-averaged number of qubits in the largest cluster $\overline{n_{\text{max}}}$ grows extensively in the connected phase while it remains $\mathcal{O}(1)$ in the disconnected phase.

This motivates us to define the following finite size scaling ansatz for the $\overline{n_{\text{max}}}$
\begin{equation}\label{eq:FSS_nmax}
    \overline{n_{\text{max}}} \sim L^{\tilde{\beta}_c/\nu_c}f[(p_{m}-p_{m}^{cc})L^{1 / \nu_c}],
\end{equation}
where the estimated critical exponents are shown in Table~\ref{tab:single_layer_cluster}.
\begin{table}[htbp]
    \centering
    \caption{Critical exponents of Connectivity Transition.
    }
    \begin{tabular}{|c|c|c|c|c|}
     \hline
       & $p_{Z \vee X}=0$ & $p_{Z \vee X}=0.25$ &  $p_{Z \vee X}=0.5$ &  $p_{Z\vee X}=1$ \\
     \hline\hline
     $p_{m}^{cc}$ & 0.66(1) & 0.61(2) & 0.61(2) & 0.66(2) \\
    \hline
     $p_{c}^{ce}$ & 0.26(2) & 0.26(1) &  0.25(3) & 0.27(2)  \\
    \hline
     $\nu_c$ & 2.5(3) & 2.3(2) &  2.3(2) & 2.0(2)  \\
     \hline
     $\tilde{\beta}_c$ & 1.80(8) & 1.63(5) & 1.66(7) & 1.48(6) \\
    \hline\hline
    \end{tabular}
    \label{tab:single_layer_cluster}
\end{table}

A close variant of this model is introduced in \cite{Vijay2020measurement}, which exhibits an exactly solvable phase transition between a fully entangled volume-law phase and a separable phase composed of finite subsystems. Because these specific unitary dynamics and projective measurements can be recast as a measurement-outcome-dependent unitary transformation, the model allows for an exact analytical treatment that maps the separability transition to mean-field percolation. Furthermore, this model serves as a compelling starting point since this entanglement transition coincides with a sharp phase transition in the computational hardness of classical simulation.

\subsection{Entanglement Transition}
\label{sec:entanglement_appendix}
To study the entanglement transition, we need to generalize the concept of entanglement entropy and the corresponding volume/area-law phase to the all-to-all model, as there is no clear notion of surface and bulk in this effective 0D system.
Therefore, we generalize the entanglement entropy to the following max-min form \cite{gullans_huse_pc_2023}
\begin{equation}\label{eq:max-min_eq}
    S_{mm,n}=\max _{C_i} \min _A S_{A}^{(n)} \leq \max _{C_i}|C_i| / 2,
\end{equation}
where $A$ is a half of a bipartition of cluster $C_i$, and $S_n$ is the $n$-th R\'enyi entropy. This max-min procedure is designed to find the most entangled cluster and then the least entangled bipartition within it, providing a robust measure that avoids being skewed by small, disconnected clusters. Definitions of entanglement which are based on averaging over bipartitions tend to yield $p_{ce} = p_{cc}$ on all-to-all circuits. To distinguish the entanglement transition from this connectivity transition, we introduce a minimization procedure over bipartitions. The generalized entanglement entropy should be extensive in the volume-law phase, and sub-extensive in the area-law phase.

\begin{figure}[htbp]
    \centering
    \includegraphics[width=3.4in]{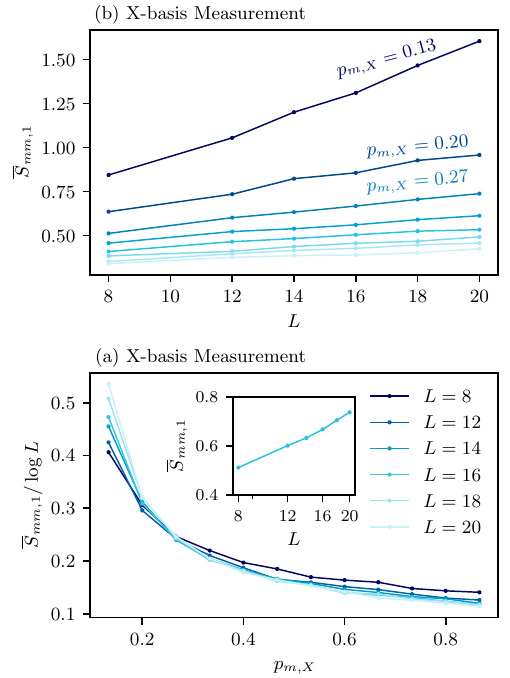}
    \caption{(a) Scaling of $\overline{S}_{mm,1}$ with $L$ for $p_{m} \in \{0.13,0.2,0.27,0.34,0.41,0.48,0.55,0.62\}$. We find the critical entanglement (ce) rate $p_{m}^{ce} = 0.27(1)$ as indicated by the inflection point. (b) Max-min entanglement entropy $\overline{S}_{mm,1}$ as a function of measurement probability $p_{m}$. $\overline{\cdots}$ denotes averaging over circuit trajectories. Data is shown for system sizes $L \in \{8,12,14,16,18,20\}$ in the single pair all-to-all model with $X$-basis measurements. The crossing suggests the location of the transition (Inset)  $\overline{S}_{mm,1}$ as a function of $L$, which exhibits log scaling at the transition.}
\label{fig:entanglement_transition_all_to_all}
\end{figure}

In Fig.~\ref{fig:entanglement_transition_all_to_all}(b), we show the max-min entanglement entropy as a function of system size. At small measurement rates, $p_{m} < p_{m}^{ce}$, $S_{mm,1}$ scales extensively in system size. At large measurement rates, $p_{m} > p_{m}^{ce}$, $S_{mm,1}$ scales sub-extensively in system size. The critical measurement rate $p_{m}^{ce}$ can thus be identified by locating the inflection point of $S_{mm,1}$ on a linear-log plot, where $S_{mm,1}$ grows logarithmically in system size, i.e.,
\begin{equation}
S_{mm,1} \sim \mathrm{log}(L).
\end{equation}
We find $p_{m}^{ce} = 0.27(1)$.
\subsection{Purification Transition}
This section provides additional data supporting the purification transition results reported in Sec.~\ref{sec:MIPT_transitions}.
We give more details on how we locate the purification transition in Section \ref{sec:MIPT_transitions}. We present the data used to numerically extract the phase boundary of the purification transition in Fig.~\ref{fig:all_to_all_purification_plot}(a). We also explain the fitting procedure used to extract the decay times $\tau$ shown in Figs.~\ref{fig:all_to_all_purification_plot}(b) and (c). We obtain the decay times $\tau$ by fitting the late-time exponential regime of $\overline{S_Q}(t)$.
\subsubsection{Fitting Procedure} \label{sec:fitting_procedure}
We do not fit all of the data, $\overline{S}_{Q}(t)$ from $t_{i}=0$ to $t_{f}=2L^{2}$ to the exponential decay ansatz $\overline{S}_{Q}(t) = Ae^{-t/\tau}.$ Instead, we fit over all possibilities of $t_{f}$ and $t_{i}$, where $t_{f}-t_{i}\geq L^{2}/8$, and $\frac{\text{err}[S_{Q}(t_{f})]}{\overline{S_{Q}(t_{f})}} < 0.35.$ The first condition ensures that an artificially small region is not being fit. The second condition ensures that the error of each datapoint is sufficiently small relative to the datapoint value. We compute $\tau$ using the region that gives the largest coefficient of determination, $R^{2}$, for the fit.

\begin{figure}[htbp]
    \centering
    \includegraphics[width=3.4in]{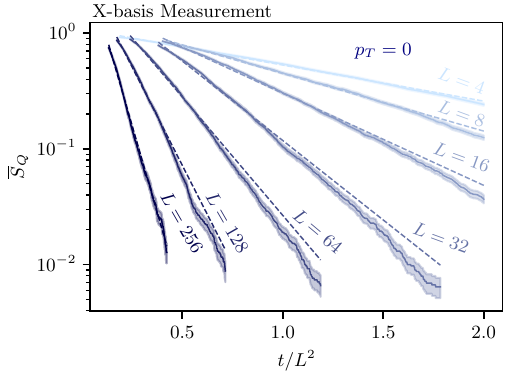}
    \caption{Late time decay of $\overline{S_{\text{Q}}}$, which is the entanglement entropy between the ancilla qubit and the system. $\overline{\cdots}$ denotes averaging over circuit trajectories. This shows the exponential decay regime used to extract the decay time $\tau$, where $\overline{S_{\text{Q}}} \sim e^{-t/\tau}$. Data is shown for the single-pair all-to-all model with $X$-basis only measurements at the critical purification rate $p_{m}^{cp}=0.26(1)$ for the purification transition.}
\label{fig:time_decay}
\end{figure}
Figures~\ref{fig:TauvL_aggregate.pdf} and ~\ref{fig:Tauvp_aggregate} show the purification transition in the single-pair all-to-all model with measurements in the $X$-basis only. In this model, where the entanglement is no longer degenerate in the rate of $T$-gates, the purification transition remains unaffected.

\subsubsection{Additional Result on Purification Transition}

\begin{figure}[htbp]
    \includegraphics[width=3.4in]{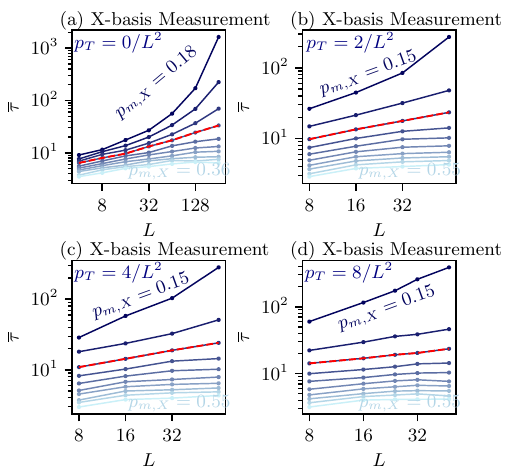}
    \caption{Dependence of the decay time $\overline{\tau}$ as a function of system size $L$, where $\overline{S_{Q}} \sim e^{-t/\tau}$ and $S_{Q}$ is the entanglement between the ancilla and the system. $\overline{\cdots}$ denotes averaging over circuit trajectories. Data is shown for the single-pair all-to-all model with $X$-basis only measurements. The measurement probabilities $p_{m}$ are obtained near the critical purification (cp) rate $p_{m}^{cp}$. We extract $p_{m}^{cp}$ and $z$, where $\tau(L,p_{m}^{cp}) \sim L^{z}$, by looking for the inflection point in this family of curves, and fitting the slope.  The probability of a $T$-gates is given as (a) $p_{T} = 0 $, (b) $p_{T} = 2 / L^{2}$, (c) $p_{T} = 4 / L^{2}$, and (d) $p_{T} = 8 / L^{2}$. The fit used to obtain z is shown as a dashed red line.
    }
    \label{fig:TauvL_aggregate.pdf}
\end{figure}

\begin{figure}[htbp]
    \centering
    \includegraphics[width=3.4in]{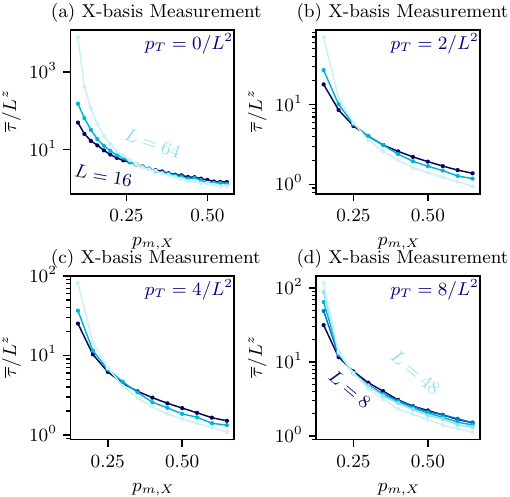}
    \caption{Dependence of the decay time $\overline{\tau}$ as a function of the measurement probability $p_{m}$, where $\overline{S_{Q}} \sim e^{-t/\tau}$ and $S_{Q}$ is the entanglement between the ancilla and the system. $\overline{\cdots}$ denotes averaging over circuit trajectories. Data is taken in the single-pair all-to-all model with $X$-basis only measurements. Here $z \approx 1/5$, where $\tau(L,p_{m}^{cp}) \sim L^{z}$, is the dynamical critical exponent fixed from the inflection point analysis at the critical purification (cp) rate, consistent with the crossing. The probability of a $T$-gates is given as (a) $p_{T} = 0 / L^{2}$, (b) $p_{T} = 2 / L^{2}$, (c) $p_{T} = 4 / L^{2}$, and (d) $p_{T} = 8 / L^{2}$.}
    \label{fig:Tauvp_aggregate}
\end{figure}

We now present data supporting the phase boundary of the purification transition (Fig.~\ref{fig:all_to_all_purification_plot}) in the single-pair all-to-all model with $X$-basis only measurements. Figures~\ref{fig:TauvL_aggregate.pdf} and~\ref{fig:Tauvp_aggregate} show the scaling behavior of the decay time $\tau$ as a function of both system size $L$ and measurement probability $p_{m}$ for different $T$-gate densities. These results provide further confirmation that the purification transition persists even in the presence of non-Clifford gates, and allow for an independent extraction of the dynamical exponent $z$ and the critical purification rate $p_m^{cp}$.

 \clearpage

\section{Statistical Properties of the LRSD Space Complexity}
\label{sec:space_appendix}
In this appendix, we provide an extended statistical analysis of the space complexity $\mathcal{N}$ to supplement the ensemble-averaged scaling results presented in Section ~\ref{sec:space_complexity}. Here, we present the full cost distributions to confirm the absence of exponentially hard trajectories, analyze the second moment of these distributions to rigorously bound the variance, and provide additional scaling data utilizing random Clifford gates to demonstrate the robustness of this efficient scaling regime across different circuit structures. Furthermore, to characterize the scaling complexity and demonstrate robustness to generic circuit structures, we completely replace the CZ gates with random Clifford gates in Fig.~\ref{fig:random_log_scaling}.

\subsection{Distribution of Simulation Costs}

Since the number of logical operators $|\{\lambda_{\ell}\}|$ directly dictates the memory and computational overhead of the LRSD algorithm [Eq.~\eqref{eq:NumberEntries}], evaluating its trajectory-resolved probability distribution, $p(|\{\lambda_{\ell}\}|)$, serves as a direct proxy for the simulation cost.

\begin{figure}[htbp]
    \centering
    \includegraphics[width=3.4in]{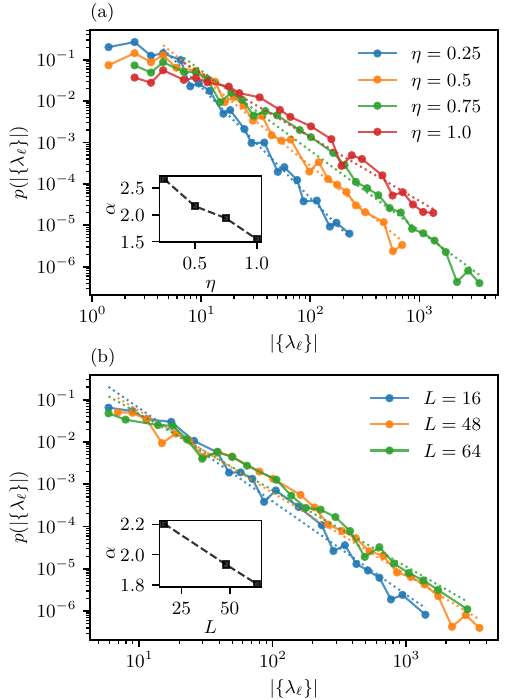}
    \caption{Probability distribution $p(|\{\lambda_{\ell}\}|)$ of the total number of logicals $|\{\lambda_{\ell}\}|$, for the single-pair all-to-all model with $X$-basis measurements and CZ gates. The distribution exhibits a power-law tail $p(x) \propto x^{-\alpha}$. Dotted lines indicate linear fits in the log-log scale to extract the decay exponent $\alpha$. (a) Data is shown for varying $T$-gate probabilities $\eta \in \{0.25, 0.5, 0.75, 1.0\}$ at a fixed system size $L=48$. The $T$-gate rate is $p_{T}=\eta/L^{\beta}$, for $\beta=1$. The inset displays the dependence of the extracted exponent $\alpha$ on $p_{T}$. (b) Scaling of the probability distribution for system sizes $L \in \{16, 48, 64\}$ at a fixed $T$-gate probability $p_{T} = 0.75$. The inset displays the scaling of the exponent $\alpha$ with $L$. For both panels, the measurement probability is fixed at $p_{m,X} = 0.6$ and the truncation cutoff is $\epsilon = 1 \times 10^{-15}$. Data was obtained by from $\sim 10^{5}$ circuit realizations.}
    \label{fig:cost_distribution}
\end{figure}

\begin{figure}[htbp]
    \centering
    \includegraphics[width=3.4in]{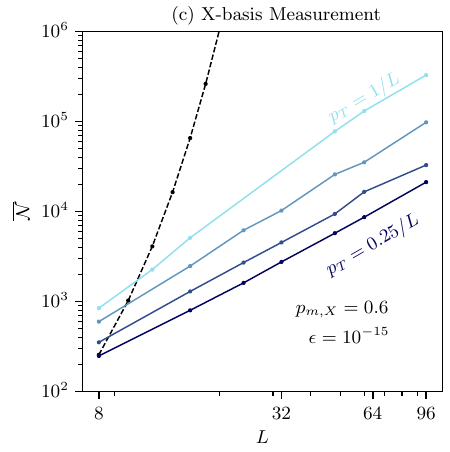}
    \caption{
        (a) Scaling of the circuit-averaged number of entries $\overline{\mathcal{N}}$ with $L$, see Eq.~\eqref{eq:NumberEntries}, which is the total number of entries required to specify the state in our algorithm. $|\{\lambda_{l}\}|$ represents the total number of logicals $\lambda_{l}$, where $\lambda_{l}$ is defined in Eq.~\eqref{eq:LSRD_decomposition}. Data is shown for the single-pair all-to-all model with $X$-basis measurements. The gates are random Clifford gates. For the state-vector simulation, $\mathcal{\overline{N}} = 2^{L}$, as shown by the black dashed line. Data is shown for $T$-gate probabilities $p_{T} \in \{1/L,0.75/L,0.5/L,0.25/L \}$. Data was obtained by averaging over $\sim 10^{3}$ circuit realizations.}
    \label{fig:random_log_scaling}
\end{figure}

In Fig.~\ref{fig:cost_distribution}, we plot $p(|\{{\lambda_{\ell}}\}|)$ for varying system sizes, $L \in \{16,48,96\}$, across a range of sub-extensive $T$-gate rates from $p_{T}=0.25/L$ to $p_{T}=1/L$. The empirical distributions demonstrate a distinct unimodal structure. As expected, as both the $T$-gate injection rate and system size increase, the mean of the distribution shifts to higher values and the variance broadens.

The probability weight on the right-hand side of the mean exhibits a power-law decay $p(|{\{\lambda_{\ell}\}|)} \sim |{\{\lambda_{\ell}\}}|^{-\alpha}.$ This is observed in for $p_{T}=0.25/L$ to $p_{T}=1/L$ at $p_{m,X}=0.6$. In contrast to regimes with exponential suppression, this behavior suggests that the likelihood of encountering highly demanding trajectories decreases more gradually. As a result, even at low injection rates, such extreme events can retain non-negligible probability weight. This observation is consistent with the presence of ``hard trajectories'' in this regime, indicating that these rare but costly realizations may play an important role in shaping the average-case sampling dynamics. Importantly, we emphasize that this feature does not restrict our claims regarding efficient simulability within the parameter regimes studied here as we reach $L=96$ despite this heavy-tailed behavior.

\FloatBarrier
\section{Magic Analysis} \label{sec:magic_analysis}
We fit the data, $\mathcal{M}(L)$ to the function $f(L)=a+bL^{\gamma}.$ For the power-law fit, we impose the constraint $\gamma > 0.$ For the area-law fit, we impose the constraint $\gamma \leq 0.$ If the best fit is $\gamma \leq 0$, then the scaling is classified as area-law. Otherwise, it is classified as power-law scaling.

To assess the quality of each fit, we compute the chi-squared fit parameter using:
\begin{equation}
\chi^2 = \sum_L \frac{ \left[ \overline{\mathcal{M}}_{\mathrm{ss}}(L) - f\left(L, a_{\mathrm{fit}}, b_{\mathrm{fit}}, c_{\mathrm{fit}}, \gamma_{\mathrm{fit}} \right) \right]^2 }{ \sigma(L)^2 }
\end{equation}
Here, $\overline{\mathcal{M}}_{\mathrm{ss}}(L)$ denotes the steady-state stabilizer nullity, and $\sigma(L)$ is the standard error of each datapoint. The degrees of freedom (dof) are the number of data points minus the number of fitted parameters.

\begin{figure}[htbp]
    \centering
    \includegraphics[width=3.4in]{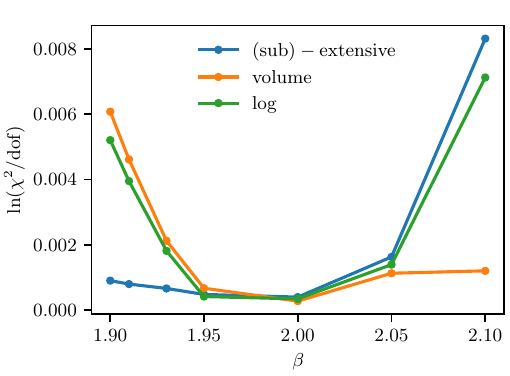}
    \caption{
        The squared residual of freedom as a function of the scaling parameter $\beta$ for the magic of a hybrid circuit with a $T$-gate density $p_{T}=\eta/L^{\beta}$, where $\eta=1$ and $p_{m}=0.5$. Data is obtained in the single-pair all-to-all model with $Z$-basis measurements, system sizes $L \in \{8,12,16,24,32,48,64,96,128,196,256\}.$
        }
    \label{fig:dof_plot}
\end{figure}

To identify a proxy for the critical $T$-gate rate $\beta_{c}$,
we plot $\chi^{2}/\mathrm{dof}$ as a function of $\beta$ for three different scaling ansatze: the (sub)-extensive and area-law power-law forms, and the logarithmic scaling function $f_{\mathrm{log}}(L) = a + b \ln L$. In Fig.~\ref{fig:dof_plot}, we plot $\chi^{2}/\mathrm{dof}$ against the scaling parameter $\beta$ for the (sub)-extensive and area law scaling fit functions. In addition, we also plot $\chi^{2}/\mathrm{dof}$ for fits of the logarithmic scaling function.

We define $E_{\mathrm{ext}}(\beta)=\chi_{\mathrm{ext}}^{2}/\mathrm{dof}$ and $E_{\mathrm{area}}(\beta)=\chi_{\mathrm{area}}^{2}/\mathrm{dof}.$ $\chi_{\mathrm{ext}}^{2}$ and $\chi_{\mathrm{area}}^{2}$ are the extensive and area law fits, respectively. We then define $\beta_{c}$ as the value at which $\mathrm{ln}[F(\beta_{c})]=0.$ To identify a transition in the measurement rate $p_{m}$ at a fixed $\beta$, we follow the same procedure.

We estimate the uncertainty $\sigma_{\beta_c}$ as
\begin{equation}
\sigma_{\beta_c} = \sqrt{\sigma_A^2 + \sigma_B^2},
\end{equation}
with two contributions:
To compute $\sigma_{A}$, we find the values $\beta^{\pm}$ where $\mathrm{ln}[F(\beta^{\pm})]=\pm1$, and then take $\sigma_{A}=(\beta^{+}-\beta^{-})/2.$
The second contribution is $\sigma_B = \left| \beta_c(L_{\max} = 256) - \beta_c(L_{\max} = 192) \right|$.
Here, $\beta_c(L_{\max}=192)$ denotes the estimate of $\beta_c$ computed by excluding data for the largest system size $L=256$.

\end{document}